\documentstyle{mn}
\newcommand{\MSUN}{M$_{\odot}$}
\newcommand{\MDOT}{\rm \dot{M}}
\newcommand{\MSUNYR}{$\rm M_{\odot}\,\rm yr^{-1}$}

\newcommand{\etal}{{\it et al.}$\;$}

\newcommand{\ltappeq}{\raisebox{-0.6ex}{$\,\stackrel
{\raisebox{-.2ex}{$\textstyle <$}}{\sim}\,$}}
\newcommand{\gtappeq}{\raisebox{-0.6ex}{$\,\stackrel
{\raisebox{-.2ex}{$\textstyle >$}}{\sim}\,$}}

\newcommand{\mdot}{\mbox{$\stackrel{.}{\textstyle M}$}}

\title{ Radiation driven winds from luminous accretion disks.}

\author[D. Proga et al.]{Daniel Proga$^{a}$, James M. Stone$^b$, and 
Janet E. Drew$^a$\\
$\rm ^a$ Imperial College of Science, Technology and Medicine, 
Blackett Laboratory, Prince Consort Road, London SW7 2BZ, UK \\
$\rm ^b$ Department of Astronomy, University of Maryland, College Park 
MD~20742, USA\\
E-mail: d.proga@ic.ac.uk, jstone@astro.umd.edu, and j.drew@ic.ac.uk}

\begin{document}
\input epsf
\maketitle

\begin{abstract}
We study the two-dimensional, time-dependent hydrodynamics of radiation-driven
winds from luminous accretion disks in which the radiation force is mediated 
primarily by spectral lines.  We assume the disk is flat, Keplerian, 
geometrically thin, and optically thick, radiating as an ensemble of 
blackbodies according to the $\alpha$-disk prescription.  The
effect of a radiant central star is included both in modifying the radial 
temperature profile of the disk, and in providing a contribution to the 
driving radiation field.  Angle-adaptive integration techniques are needed
to achieve an accurate representation of the driving force near the surface 
of the disk.  Our hydrodynamic calculations use non-uniform grids to resolve 
both the subsonic acceleration zone near the disk, and the large-scale global 
structure of the supersonic wind.

We find that line-driven disk winds are produced only when the effective 
luminosity of the disk (i.e. the luminosity of the disk times the maximum 
value of the force multiplier associated with the line-driving force) exceeds 
the Eddington limit.   If the dominant contribution to the total radiation 
field comes from the disk, then we find the outflow is intrinsically 
unsteady and characterised by large amplitude velocity and density 
fluctuations.  Both infall and outflow can occur in different regions of the 
wind at the same time.  The cause of this behaviour is the difference in the
variation with height of the vertical components of gravity and
radiation force: the former increases while the latter is nearly
constant.  On the other hand, if the total luminosity of the system is
dominated by the central star, then the outflow is steady.  In either
case, we find the two-dimensional structure of the wind consists of a
dense, slow outflow, typically confined to angles within $\sim$45
degrees of the equatorial plane, that is bounded on the polar side by a
high-velocity, lower density stream.  The flow geometry is controlled
largely by the geometry of the radiation field -- a brighter disk/star
produces a more polar/equatorial wind.  Global properties such
as the total mass loss rate and terminal velocity depend more on the
system luminosity and are insensitive to geometry.  The mass loss rate
is a strong function of the effective Eddington luminosity; less than
one there is virtually no wind at all, whereas above one the mass loss
rate in the wind scales with the effective Eddington luminosity as a
power law with index 1.5.  Matter is fed into the fast wind from within
a few stellar radii of the central star.

Our solutions agree qualitatively with the kinematics of outflows in 
CV systems inferred from spectroscopic observations.  We predict that low 
luminosity systems may display unsteady behavior in wind-formed spectral
lines.  Our study also has application to winds from active galactic nuclei 
and from high mass YSOs.

\end{abstract}


\section{Introduction}

    Radiation-pressure driven wind models for main sequence and evolved
OB stars developed over the past two decades have proven enormously
successful in accounting for the gross properties of such outflows.
The essential concept underpinning these models is that momentum is
extracted most efficiently from the radiation field via line
opacity (Lucy \& Solomon 1970; Castor, Abbott \& Klein 1975, hereafter
CAK).  With the inclusion of lines, CAK showed that the effective
radiation force can be increased by several orders of magnitude above
that due to electron-scattering alone, thus facilitating mass loss even
from stars radiating at around 0.1\% of their Eddington limit.  A
decade later it had become clear that the model was well able to
predict time-averaged mass loss rates and terminal velocities in
agreement with empirical estimates (Friend \& Abbott 1986).  More
recently still, the goal has been to obtain a deeper understanding of
instabilities inherent in the line-driving mechanism (e.g. Owocki,
Castor \& Rybicki 1988; Puls, Owocki \& Fullerton 1993).  These
impressive achievements have all been set in the context of
one-dimensional (spherical) geometry.

    At the same time, accretion disks have come to be accepted as
important components in a wide variety of astrophysical settings.
These too can produce intense radiation fields at effective radiation
temperatures comparable with those of OB stars.  As an example, such
disks are inferred from observations to be present in
high state cataclysmic variables (CVs, see Warner 1995).  Moreover, 
observations also show these same systems give rise to very high velocity 
winds that most probably emanate from deep within the gravitational potential 
in the vicinity of the accreting star (see Drew 1997).
Another example of outflow associated with a luminous disk may be
provided by higher mass YSOs (the BN-type objects and Herbig Be stars,
see Mundt \& Ray 1994 and references therein).  In fact, in these objects it 
would not even be necessary
to sustain a high mass accretion rate to achieve an intense radiation
field -- as one is already produced by the young OB star. Finally,
shifting up the luminosity scale many orders of magnitude, the
accretion disk model is now `standard' for active galactic nuclei, and
in this context as well, there is direct evidence of fast outflow in
the broad absorption line (or BAL) QSOs (e.g. Weymann et al. 1991).

    Given the success of one-dimensional models of line-driven winds from
hot stars, it is natural to ask: what is the nature of line-driven winds
from a star plus luminous accretion disk?  In practice, calculating models
for winds in disk systems is complex because of the intrinsically 
two-dimensional, axisymmetric character of the problem.

    To render this problem amenable to analytic solution, previous
studies have generally found it necessary to introduce simplifying or ad hoc 
assumptions.  For example, in seeking a steady state accretion disk wind 
solution, Vitello \& Shlosman (1988) found it necessary to enforce a 
radiation force term that increased with height above the disk and required
a very particular variation in the ionization state of the gas -- a matter 
that gave them cause for concern.  More recently an analytic disk wind model 
has been designed by Murray \etal (1995) specifically for AGN.  In order to 
simplify the problem they introduced heuristic assumptions which allowed the 
equations of motion in the radial and polar angle directions to be solved 
separately.  Unfortunately, the outcome of their calculations depends in a 
basic way on these assumptions.

    In this study, we face the multi-dimensional character of the disk wind 
problem directly, by adopting numerical methods to solve the dynamical 
equations from first principles. The first numerical treatment of the problem 
was formulated as long ago as 1980 (Icke 1980, see also Icke 1981).  Icke 
(1980) set up what, in terms of today's computing power, would now be 
regarded as a very modest calculation aimed at determining the character of 
an outflow from a disk driven by continuum (electron-scattering) radiation 
pressure only.  In 1981 he incorporated rotation into his treatment and
obtained results that we have found to be broadly comparable with our own as 
a partial test of our independent formulation.  Like Icke, we choose a 
two-dimensional computational domain in which the central accreting object 
and inner disk are well-resolved.  In our view this is important because of the
evidence that disk winds in cataclysmic variables, a natural first test-bed 
for our results, originate close to the white dwarf.  Recently Pereyra, 
Kallman \& Blondin (1997, see also Pereyra 1997) have also presented numerical
calculations of the two-dimensional structure of CV disk winds, albeit at a 
resolution too coarse to capture the inner disk structure or the subsonic 
portion of the outflow.  Here, we use non-uniform meshes at high resolution 
to capture the structure of the wind in both the subsonic and supersonic 
regimes.  Moreover, we employ a carefully-tested adaptive numerical 
integration technique to compute the line force directly within the Sobolov 
approximation.  We make no restrictive geometric assumptions with regard to
the flux integrals involved and show, on holding the force multiplier
constant, that the line-driving force should be constant near the disk.

    The original motivation for these calculations was to obtain
self-consistent dynamical models for CV winds that remove the need to
apply ad hoc kinematic structures in modelling observed ultraviolet
spectral line profiles.  Recent high quality observations obtained
with the Hubble Space Telescope (e.g. Mason et al. 1995) have shown that 
the kinematic models designed for IUE data are already inadequate (Knigge
\& Drew 1997), and serve to emphasise the need for realistic rather than
simplistic models.

    Our models incorporate a star, for which we adopt the mass and
radius appropriate to a white dwarf, and a geometrically thin
accretion disk that is a source both of radiation and mass.  In our
formulation of the problem, the velocity field has three components
that are functions of two spatial coordinates (sometimes referred to
as a 2.5D formulation). For this reason, exact implementation of a force
multiplier into the radiation force on spectral lines is formidable and 
some level of approximation required.  We describe the formal solution of 
the problem and our approximations in section 2, with most of the analytic 
details given in appendices.  We have developed numerical methods to compute 
both the radiation field from, and the radiation driving force, associated with
the disk.  These are coupled to the hydrodynamical code ZEUS-2D (Stone
\& Norman 1992).  We describe our numerical methods and tests
in section 3. Our particular interests are to explore the impact upon
the mass loss rate and outflow geometry of (i) varying the system
luminosity and (ii) varying the radiation field geometry by changing
the relative contributions of the central star and disk radiation
fields.  Crudely speaking, we find that the mass loss rate is an
extremely strong function of luminosity, while the outflow geometry
and its temporal behaviour are controlled by the radiation field's
geometry.  Some of our results have already been presented in a short
communication (Proga, Drew \& Stone 1997).  A full description of our
results is given in section 4.  These are discussed together with their
likely relevance and perceived limitations in section 5.
The paper ends, in section 6, with our conclusions.

\section{The representation of the line driving force}

To calculate the radiation force from a disk we need first to specify the
disk's geometry and its radiation field.  Consistent with existing
conventions, we assume a flat, Keplerian, geometrically-thin and 
optically-thick disk.  We calculate the disk
radiation field from the surface brightness of the so-called 
$\alpha$-disk (Shakura \& Sunyaev 1973).  For models including a radiant
central star, we take into account its irradiation of the disk,
and assume that the disk re-emits all absorbed
energy locally as a black body.  The irradiation increases the disk
temperature primarily in the inner part of the disk. In the presence
of a luminous central star (CS), the disk temperature profile is altered
such that the temperature decreases monotonically with radius, whereas 
a pure $\alpha$-disk is characterised by a temperature maximum at 1.36 
stellar radii.  See Appendix~A for further details.

We approximate the radiative line force by means of the formalism 
introduced by CAK.  A key assumption in their method is that the Sobolev
approximation is valid, i.e. the radiation force exerted as a result of 
absorption of radiation coming from the central radiant object along a 
direction $\hat{n}$ depends mainly on the velocity gradient along $\hat{n}$, 
at the point of absorption.  CAK showed that for a spherically-expanding 
flow the radiation force due to a large ensemble of spectral lines can be 
expressed by
\begin{equation}
F^{rad,l} = F^{rad,e} M(t),
\end{equation}
where $F^{rad,e}=\frac{\sigma_e}{c}{\cal F}_{\ast}$, is the radiation
force due to electron scattering, and ${\cal F}_\ast$ is the frequency
integrated flux from the star.  The quantity $M$, called the force 
multiplier, represents the
increase in radiation force over the pure electron scattering case
when lines are included.  In the Sobolev approximation, it is
a function of the optical depth parameter
\begin{equation}
t~=~ \sigma_e \rho v_{th} \left|\frac{dv}{dr}\right|^{-1},
\end{equation}
where $\rho$ is the density, $v_{th}$ is the thermal velocity, and
$\frac{dv}{dr}$ is the velocity gradient along the radial direction
(the only non-zero component in spherical symmetry).  CAK found a fit 
to their numerical results for the force multiplier, such that
\begin{equation}
M(t)~=~k t^{-\alpha},
\end{equation}
where $k$ and $\alpha$ are parameters of the fit.  They also showed
that this expression for the force multiplier could be reproduced by
assuming a simple statistical model of the distribution of lines in
strength and frequency. This statistical model indicates that $k$ is
proportional to the total number of lines involved and $\alpha$ is the
ratio of optically-thick to optically-thin lines.

Subsequently the CAK formalism to approximate the radiation force due
to lines has undergone a number of refinements (e.g., Abbott 1982; Pauldrach,
Puls \& Kudritzki 1986; Stevens \& Kallman 1990, Gayley 1995). Many of
the modifications are specific to 1D spherically-symmetric stellar winds.  
In this work we find the refinement introduced by Owocki, Castor and
Rybicki (1988, OCR hereafter) is important.  The formal specification
of the force multiplier used by CAK allows an unlimited increase of the 
radiation line force with decreasing $t$, which is clearly unphysical (a point
remarked upon by CAK).  Instead one expects the 
force multiplier to saturate at some maximum value for very low $t$ as all 
lines, including the most optically thick, contribute to the radiation force: 
a further decrease of $t$ does not 'activate' any more lines.  For example, 
this saturation can be seen in the radiation force calculations of Abbott 
(1982) in which he accounted for lines of the first to sixth stages of 
ionization of the elements H-Zn (see his figure~3).  Generally he confirmed 
the CAK results. However his results showed that for $t \ltappeq 10^{-7}$, 
M(t) falls away from the CAK approximation (equation 3) as the force 
multiplier becomes less sensitive to $t$, the optical depth parameter.  
OCR considered this problem in terms of a line strength distribution.  They 
modified the simple CAK statistical model by cutting off the maximum line 
strength and thereby limiting the effect of very strong lines:
\begin{equation}
M(t)~=~k t^{-\alpha}~ 
\left[ \frac{(1+\tau_{max})^{(1-\alpha)}-1} {\tau_{max}^{(1-\alpha)}} \right]
\end{equation}
where $\tau_{max}=t\eta_{max}$ and $\eta_{max}$ is a parameter determining
the maximum value, $M_{max}$ achieved for the force multiplier.  
Equation (4) shows the following limiting behaviour:
\begin{eqnarray}
\lim_{\tau_{max} \rightarrow \infty}~M(t) & = & k t^{-\alpha} \\
\lim_{\tau_{max} \rightarrow 0}~M(t) & = & M_{max},
\end{eqnarray} 
where $M_{max} = k (1-\alpha)\eta_{max}^\alpha$.
The maximum value of the force multiplier is in reality a function
of physical parameters of the wind and radiation field.  In a number of
studies (CAK, Abbott 1982, Stevens \& Kallman 1990, Gayley 1995) it has 
been shown that $M_{max}$ is of the order of a few thousand.

To adapt the CAK formalism designed for OB stars to the disk wind case
we need to accommodate two essential differences: (1) a stellar wind
can be well approximated by a 1D radial flow while a disk wind is in
general a 3D flow; (2) the stellar radiation field is spherically
symmetric while the disk radiation is axially symmetric, as a
consequence of the disk geometry and the non-uniform disk intensity.
First we consider the 3D nature of the flow.  Rybicki \& Hummer (1978,
1983) generalized the Sobolev method to the 3D case in which the flow
velocity along a line of sight is not necessarily monotonic.  In such
a case, radiative coupling between distant parts of the flow must be
taken into account. At this stage, where the flow properties are not
known, we do not consider this effect.  However, a further consequence
of generalising the Sobolev method to 3 dimensions is that the Sobolev 
optical depth's dependence on the photon line-of-flight velocity gradient,
$\left|dv_l/dl\right|$, may in practice become a complicated function of the 
velocity, velocity derivatives and position.  
In the generalized Sobolev method
\begin{equation}
t~=~t(\hat{n},{\bf v})=\frac{\sigma_e \rho v_{th}}{ \left| dv_l/dl \right|},
\end{equation}
where the velocity gradient along the line of sight  
may be written
\begin{equation}
\frac{dv_l}{dl}=Q~\equiv~ \sum_{i,j}\frac{1}{2}\left(\frac{\delta v_i}{\delta r_j}
+\frac{\delta v_j}{\delta r_i}\right)n_in_j=\sum_{i,j}e_{ij}n_in_j
\end{equation}
and $e_{ij}$ is the symmetric rate-of-strain tensor.
Expressions for the components of $e_{ij}$ in the spherical polar coordinate
system are given in Batchelor (1967).  The complexity of the disk radiation 
field, together with the generalized optical depth parameter $t$, mean that 
the radiation line force at a given location due to the total radiating 
surface becomes a complicated integral in which the dependences on
geometry, the radiation field and local optical depth are no longer
separable.

For a 2.5D wind, we evaluate the disk radiation force in three
steps.  First, we calculate the radiation flux due to a surface
element of the disk at a point above the disk, $d{\cal F}_D$ (see Appendix A).
Then we calculate the radiation force exerted by this flux via electron
scattering, $d{\bf F}^{rad,e}_D~=~\hat{n}\frac{\sigma_e}{c} d{\cal
F}_D$, and via an ensemble of lines 
$d{\bf F}^{rad,l}_D~=~\hat{n}\frac{\sigma_e}{c}d{\cal F}_D M(t)$ (see Appendix
B).  Finally we integrate $d{\bf F}^{rad,e}_D$ and 
$d{\bf F}^{rad,l}_D$ over the total disk surface visible 
at the point in question (i.e., we exclude the 
disk region shadowed by the CS, see Appendix B).  Our calculations of the 
radiation force contributed by the CS assume that it radiates as a
blackbody at a fixed temperature and without any limb-darkening.  We 
express the CS luminosity in alpha disk luminosity units $x = L_\ast/L_D$. 
The method of calculation of the radiation force from the CS is the same as for
the disk, and  takes into account disk occultation of the CS.  In
practice, evaluating the generalised optical depth (equation 7) using all
terms in $Q$ is computationally prohibitive.  Instead, we keep only the
dominant terms (see section 3.2 and Appendix C).

\section{Numerical methods }

\subsection{ Hydrodynamics }

We use the 2D Eulerian finite difference code ZEUS-2D (Stone \& Norman 
1992) to calculate the wind structure.  We have extended the code to
include radiation forces due to electron scattering and line
driving, i.e. we solve:
\begin{equation}
   \frac{D\rho}{Dt} + \rho \nabla \cdot {\bf v} = 0,
\end{equation}
\begin{equation}
   \rho \frac{D{\bf v}}{Dt} = - \nabla (\rho c_s^2) + \rho {\bf g}
 + \rho {\bf F}^{rad}
\end{equation}
where ${\bf g}$ is the gravitational acceleration of the CS, and
${\bf F}^{rad}$ is the total radiation force per unit mass.  We describe how
${\bf F}^{rad}$ is evaluated numerically in section 3.2.  The gas in the wind
is taken to be isothermal with a sound speed $c_s$.

Our calculations are performed in spherical polar coordinates $(r,
\theta, \phi)$ with $r=0$ at the centre of the accreting
CS.  We assume axial symmetry about the rotation axis of the accretion disk 
($\theta=0^o$).  Thus we assume that all quantities are invariant in $\phi$.  
Our standard computational domain is defined to occupy the radial range 
$r_\ast \leq r \leq 10r_\ast$, where $r_\ast$ is the CS radius, and angular 
range $0 \leq \theta \leq 90^o$ (in section 4.1 we compare the results to 
a solution computed on a grid ten times larger in radial extent, i.e. 
$r_\ast \leq r \leq 100r_\ast$).

The $r-\theta$ domain is discretized into zones.  The gridding needs to
be such as to ensure that the subsonic portion of the model outflow is
sampled by at least a few grid points in both $r$ and $\theta$.  This
requirement and the nature of the problem combine to demand an
increasingly fine mesh toward the disk plane: here the density declines
dramatically with height, and, moreover, the velocity in the wind
increases rapidly.  Our standard numerical resolution consists of 100
zones in each of the $r$ and $\theta$ directions, with fixed zone size
ratios, $dr_{k+1}/dr_{k}=d\theta_{l}/d\theta_{l+1} =1.05$.  The
smallest radial grid zone has dimension $\Delta r_1=3.445\times 10^{-3}
r_\ast$ at $r = r_\ast$, while the smallest angular grid zone is
$\Delta \theta = 0.^o029$ near $\theta = 90^o$.  Gridding in this
manner ensures good spatial resolution close to the radiating surfaces
of the disk plane and the CS.  In addition, to check that our solutions
are resolved, we have computed the evolution of two of our models (our
fiducial unsteady and steady wind cases -- see sections 4.1 and 4.2
respectively) at twice this resolution, i.e. using 200 zones in each of
the $r$ and $\theta$ directions, with 
$dr_{k+1}/dr_{k}=d\theta_{l}/d\theta_{l+1} =1.025$.  We find that the
global properties of the solutions (such as the terminal velocity and
mass loss rate) differ by no more than 10\% between the high and standard
resolution models.

The boundary conditions are specified as follows. At
$\theta=0$, we apply an axis-of-symmetry boundary condition.  For
the outer radial boundary, we apply an outflow boundary 
condition.  For the inner radial boundary $r=r_{\ast}$
and for $\theta=90^o$, we apply reflecting boundary conditions.

The initial density profile is given by the condition of hydrostatic 
equilibrium in the latitudinal direction
\begin{equation}
\rho = \rho_{0} \exp \left (- \frac{G M_\ast}{2 c_{s}^2 r 
\tan^2 \theta} \right).
\end{equation}
where $\rho_0$ is the density in the first grid zone above the equatorial 
plane.  Physically, $\rho_{0}$ is analogous to the density in the photosphere 
of the disk at the base of the wind.  The interior of the disk itself is 
treated as negligibly thin and is excluded from the models (for a disk 
temperature of $10^{4}$~K at $r=2 r_\ast$, the disk scale height $H$ is 
$H/r_\ast \sim 10^{-3}$).  The value chosen for $\rho_{0}$ is arbitrary.  
Typically we choose $\rho_{0} = 10^{-9} \rm g~cm^{-3} $.  As discussed in 
section 4.4, we find the gross properties of the winds are unaffected by the 
value of $\rho_0$ provided it is large enough that the acceleration of the 
wind up to the sonic point is resolved with at least a few grid points.
We set a lower limit to the density on the grid of $\rho_{min}=10^{-22}~\rm 
g cm^{-3}$ and enforce it at all times in all models.

    For the initial velocity field, we adopt $v_r =  v_\theta = 0$, and
$v_\phi =\sqrt\frac{GM_\ast}{r}/ \sin \theta$.  We find only the
initial, transient evolution of the wind is affected by this choice of
initial velocity conditions.  At late times, we find all solutions for
the same model parameter values have the same time-averaged properties
regardless of the initial conditions.  In order to represent steady
conditions in the photosphere at the base of the wind, during the
evolution of each model we continue to apply the constraints that in
the first zone above the equatorial plane the radial velocity $v_r=0$,
the rotational velocity $v_\phi$ remains Keplerian, and the density is
fixed at $\rho = \rho_0$ at all times.  We have found that this
technique when applied to calculations of spherically symmetric
line-driven winds from stars, produces a solution that relaxes to the
appropriate CAK solution within a few dynamical crossing times.

\subsection{The radiation force }

The integrals (Appendix B) that express the radiation force have to be 
evaluated numerically.  To include the contribution from all radiating 
surface elements properly, the size of each surface element seen from a point 
W in the disk wind has to be in proportion to the distance from W to this 
surface element.  In other words, the sampling should be uniform in apparent 
solid angle at W.  A more subtle issue is that the vector character of the
radiation force means that cancellation of opposing contributions to its
net value can occur (e.g., in the radial component for points close to the
disk plane).  Care must therefore be taken in the numerical scheme that 
inappropriate gridding does not misrepresent such cancellations and thereby 
introduces spurious, fluctuating force terms.  To satisfy both these important 
requirements we calculate the radiative force using grids of radiating 
surface elements that are adapted to every point W in the wind.  The
need to use adaptive integration methods is demonstrated by the tests 
discussed in section 3.3.

The integration scheme applied to the disk component of the radiation force
uses a 2D version of Gaussian quadrature such that the number of quadrature 
points increases with increasing $\theta$.
Because of foreshortening, the 
integrands (equations B3 and C7) are strong functions of the position on 
the disk plane for $\theta$ close to $90^o$ -- they reach a maximum close to 
the point $D_m$, ($r_m, 90^o, 0^o$), wherein $r_m$ is defined relative to $r$ 
by $n_r = 0$ for $\theta=90^o$ and $\phi=0^o$ (see Appendix~A). Note 
that for $\theta \ltappeq 90^o$, the point $D_m$ is very close to the point 
on the disk for which the size of a surface element has to be the smallest 
(i.e. $D_m$ lies close to the projection of W onto the disk plane).  These 
properties of the integrands allow us to evaluate the radiation force 
accurately using a manageable number of grid points because the finest 
resolution on the disk plane is necessary mainly close to the point $D_m$.  
For the region far from $D_m$, we can afford less dense coverage of the 
disk surface.  

In evaluating the integrals involved in the disk component of the radiation 
force (equations B3 and C7), we generally break the inner radial integral 
over ($r_i, r_o$) into two sub-integrals, spanning ($r_i, r_m$) and 
($r_m, r_o$).  For cases where the point $D_m$ falls within the inner, or 
beyond the outer, edge of the disk just the one integral over ($r_i, r_o$) is 
performed.  The discretization of the disk surface uses 128, 256, 512, 1024 
radial quadrature points for $\theta \le 87.^o547$, $87.^o547 < \theta \le 
89.^o518$, $89.^o518 < \theta \le 89.^o788$, and  $\theta > 89.^o788$, 
respectively.  We calculate the outer azimuthal integral over the angular 
range ($0,\pi$) with 128 and 512 quadrature points for $\theta \le 89.^o002$ 
and $\theta > 89.^o002$, respectively. 

For the CS radiation force (equations B9 and C3), we use a 2D
version of the trapezium method with the number of quadrature points
increasing with decreasing $r$.  In the polar direction, the stellar
surface is divided up into 5000 and 500 grid points for $r \le 1.52
r_\ast$ and  $r > 1.52 r_\ast$ respectively.  In the azimuthal
direction, the resolution is 101 grid points for all $r$.

The evaluation of the line radiation force integrals is a major element in the 
computational cost of these simulations.  The situation is exacerbated by the 
need to sample the subsonic part of the wind, close to the disk plane, 
reasonably well both in $\theta$ and $r$.  Indeed we find we cannot afford a 
full recalculation of the radiation line force for all locations, at every 
time step.  Therefore we have to seek a working compromise between regular 
updating of the radiation line force and maintaining its accuracy in space. 
This suggests two contrasting ways of approximating the line force 
calculations: (1) the wind streamlines can be assumed to favour particular 
approximations to the velocity gradient that allow the radiation line force 
to be updated every time step;  (2) the wind velocity can be assumed not to 
change rapidly with time at a given location, with the implication the line 
force need only be updated after some time interval taking into account the 
exact velocity vector.  
   
    The first approach amounts to a simplification of $Q$ (equation 8) that 
enables the time-varying velocity gradient to be taken out of the integration 
over the radiating surfaces.  For present purposes, this is our preferred
option and our particular implementation of it is presented in Appendix C.
In physical terms, our treatment is equivalent to assuming that the 
transmission of disk light to points W in the wind will be controlled by the 
velocity shear in the vertical direction, and that the CS light has mainly to 
propagate through radial velocity shear.  In specifying $Q$, this allows us 
to consider just $dv_z/dz$ (where $z$ is height above the disk plane) when 
computing the disk's contribution to the force and just $dv_r/dr$ when 
calculating the CS component.  This treatment should work well for the CS 
component throughout most of the computational domain and it should also be 
adequate for the disk component except, perhaps, close to the disk plane 
where neglect of the radial and azimuthal derivatives of the rotational 
component of motion ($v_{\phi}$) will result in underestimation of the line 
force (i.e. inclusion of these terms would enhance $Q$ for $\theta 
\ltappeq 90^o$).  This neglect has less impact on the CS force 
component because foreshortening of the spherical stellar surface acts to 
reduce the weight of contributions from the larger angle, higher shear lines 
of sight with respect to the more nearly radial, low shear directions.  

    Within the present context in which unsteady flow has to be described, 
the second numerical approach of only updating the line force after a number 
of time steps requires more computational time.  It does become both viable 
and accurate where the outflow achieves a steady state rather than a 
configuration that is steady only in a time-averaged sense.  We are able to 
assess the impact of approximating $Q$ upon our simulations in such cases.  
We return to this issue in section 4.4.

\subsection{Tests of the numerical integration of the radiation force } 

    The hydrodynamical algorithms implemented in the ZEUS-2D code have
been extensively tested already (Stone \& Norman 1992), thus here
the primarily concern is with tests of the integration of the radiative
force term in the code.  As a first check, we have tested that our
numerical methods are able to reproduce the appropriate CAK wind
solutions for spherically symmetric radiation fields.  On the other
hand, tests of fully multidimensional radiation fields are more
difficult to construct.  Thus our tests are restricted to the
asymptotic behaviour of the electron-scattering component of the
radiation force in a few cases where we know the analytical solution.

    As $\theta$ approaches $90^o$, we can treat the disk as an isothermal 
infinite plane.   Therefore as $\theta \rightarrow 90^o$ we expect the  
radial component of ${\bf F}^{rad,e}_{D}$ to approach zero, while the 
latitudinal component of ${\bf F}^{rad,e}_{D}$ becomes $\theta$-independent 
and equal to that exerted by the local disk flux.  For high $r$ and low 
$\theta$, the CS can be approximated by a point source.  Thus the radial 
component of ${\bf F}^{rad,e}_{\ast}$ should  decrease there like $r^{-2}$ 
and the latitudinal component of the radiation force should vanish.
The radiation force calculated according to the numerical scheme
described in \S3.2 is a smooth function of $r$ and $\theta$, and agrees
with the above asymptotic solutions to within 5\% . The discrepancy is
attributable to fact that our disk is not isothermal -- even for points
close to the disk, regions of different temperature contribute to the
force.

\begin{figure}
\begin{picture}(80,170)
\put(0,0){\includegraphics{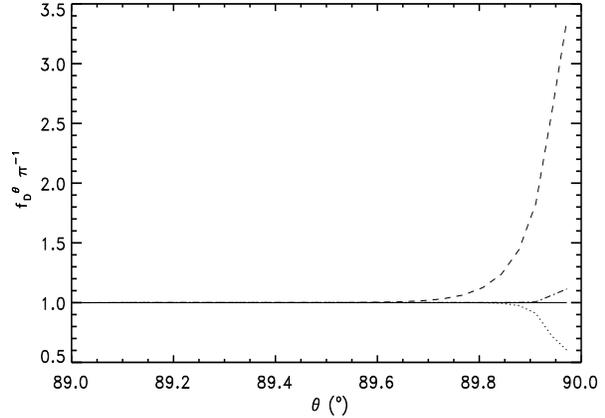}}
\end{picture}
\caption{
Numerical estimates of the vertical component of the geometric
integral entering the radiation force for the isothermal disk case (equation 
12).  The results, normalised to the expected value of $\pi$, obtained for 
different numbers of grid points are shown. The solid line is obtained for
1024 adaptive grid elements, while the dashed-dotted, dotted, and dashed 
lines correspond respectively to estimates for 512, 256 and 128 quadrature 
points.
}  
\end{figure}

In the case that the disk is isothermal and without any CS irradiation
($x=0$), the $\theta$-component of equation B3 can be written as
\begin{eqnarray}
f^{\theta}_{D}({r'},\theta) & = & \int_{\phi_l}^{\phi_u} \int_{r'_i}^{r'_0} 
\frac{r'_D~\cos~\theta~\cos~\phi_D}{ d'_D} \nonumber \\
  &  &  \frac{r'~\cos~\theta} {{d'_D}^3}  r'_Ddr'_D d\phi_D.
\end{eqnarray}
For points very close to the disk plane and far from the disk edges
the solution of equation 12 should be equal to $\pi$.  Figure~1 illustrates 
how the numerical estimate of equation~(12) depends on the number of grid 
points for $r=2 r_\ast$.  As few as 128 quadrature points produces a
satisfactory integral for points W at $\theta \ltappeq 89^o.5$. 
For higher $\theta$ however, only solutions based on 1024 quadrature points 
give an accurate estimate.  Recall that in order to resolve the subsonic 
portion of the acceleration zone, we adopt a nonuniform grid with the 
smallest zones having an angular extent of only $0.^o029$.  Hence, there are 
many grid points in the region $89^o.5 \leq \theta \leq 90.^o0$ where the use 
of densely-sampled adaptive quadratures is critical. 

Before settling on the discretization described in \S3.2  we tried using the 
trapezium method to integrate the radiation force due to the disk. To achieve 
as good agreement with the asymptotic solutions as obtained using the
Gaussian scheme of \S3.2, the trapezium method requires at least 2
orders of magnitude more grid points.  We have also tried a modified
version of the trapezium method using an exponential distribution of
grid points around the point $D_m$ -- although this method gave better
results than uniformly distributed grid points, it remained inferior to
Gaussian quadrature.

We have also experimented with the number of grid points used in calculating 
the radiation force due to the CS.   This calculation is less
demanding than that for the disk component.  A reasonable number of grid 
points (as specified in \S3.2) gives satisfactory agreement with the 
analytical solution for the asymptotic case.

\section{  Results}

Our numerical models are specified by a number of parameters.
The CS is specified by its mass $M_{\ast}$ and radius $r_\ast$.
In all our calculations we assume $M_{\ast} = 0.6$~\MSUN which  yields 
the CS radius $r_{\ast} = 8.7\times10^{8}$~cm using the mass-radius 
relation for CO white dwarfs due to Hamada \& Salpeter (1961).   
The accretion disk is characterized by the mass accretion rate through it, 
$\MDOT_{a}$ (which we treat as a free parameter), and by the sound 
speed $c_s$ (fixed at 14 $\rm km~s^{-1}$).  Finally, the line-driving 
force is determined by the force multiplier parameters, $k$, $\alpha$, 
and $M_{max}$, the maximum value allowed for the force
multiplier, and by the thermal speed $v_{th}$ which sets the line widths.  
As a starting point, we adopt typical OB star values for the force multiplier 
parameters, $k$ and $\alpha$ (i.e. $k=0.2$, $\alpha = 0.6$, see Gayley 1995), 
and subsequently vary $\alpha$.  Tables 1 and 2 specify the parameter values 
of all the models discussed in sections 4.1, 4.2, and 4.3 below.

\begin{table}
\footnotesize
\begin{center}
\caption{ Full list of model parameters. The second column
shows the parameters for model 2, while the third
column indicates the parameter range explored for those parameters
which we have varied.}
\begin{tabular}{l l  l  } \\ 

\hline
  &  &  \\
Parameter & Value & Range \\
  &  &  \\
\hline 
$M_\ast$ &   0.6~M$_{\odot}$  &  \\
$r_\ast$ & $8.7 \times 10^8$ cm  &   \\
$c_s$    & 14 km s$^{-1}$  &  \\
$v_{th}$ & 0.3 $c_s$   &  \\
$\alpha$       & 0.6  & 0.4 -- 0.8 \\
$k$       & 0.2  &  \\
$M_{max}$       & 4400 &    \\
$\rho_{0}$    & $10^{-9}$~g~cm$^{-3}$ &    \\
$\rho_{min}$   & $10^{-22}$~g~cm$^{-3}$ &    \\
$\mdot_{a}$ & $10^{-8}$~M$_{\odot}$~yr$^{-1}$  & $\pi \times 10^{-9}$ \\
  &  & -- $\pi \times 10^{-7}$~M$_{\odot}$~yr$^{-1}$ \\
$x$       &  0 & 0 -- 10 \\
$r_i $     & 1~$r_\ast$ &    \\
$r_o $      & 10~$r_\ast$ &   \\
\hline
\end{tabular}
\end{center}
\normalsize
\end{table}

\begin{table*}
\footnotesize
\begin{center}
\caption{ Summary of parameter survey.}
\begin{tabular}{l c r c c r l  } \\ \hline 
         &                &   &                       &      \\
Run & $\alpha$ & $\MDOT_a$  & $x$  & $\MDOT_w$   &  $v_r(10 r_\ast)$ & comments \\ 
 &   & (M$_{\odot}$ yr$^{-1}$) &  & (M$_{\odot}$ yr$^{-1}$) & $(\rm km~s^{-1})$ &  \\ \hline  

         &                &   &                       &    \\
1  & 0.6     &  $\pi \times 10^{-9}$ & 0 &                       &      &  no supersonic outflow  \\
2  & 0.6     &  $10^{-8}    $ & 0 &$ 4.8\times10^{-14}$   &  900 & fiducial complex wind (see \S4.1)\\
3  & 0.6     &  $\pi \times 10^{-8}$ & 0 &$ 4.7\times10^{-12}$   & 3500 & complex wind  \\
4  & 0.6     &  $  10^{-7}$   & 0 &$ 4.0\times10^{-11}$   & 4500 & complex wind\\
5  & 0.6     &  $\pi \times 10^{-7}$ & 0 &$ 3.1\times10^{-10}$   & 7500 & complex wind\\ 

         &                &   &            &      \\
6  & 0.6     &  $\pi \times 10^{-9}$ & 1 &                      &  & no supersonic outflow\\
7  & 0.6     &  $10^{-8}    $ & 1 &$1.3\times10^{-12}$   & 2000 & complex wind\\
8  & 0.6     &  $\pi \times 10^{-8}$ & 1 &$1.2\times10^{-11}$   & 3500 & fiducial steady outflow (see \S4.2) \\
9  & 0.6     &  $  10^{-7}$   & 1 &$1.5\times10^{-10}$   & 6000 & steady state\\

         &                &   &            &      \\
10 & 0.6     &  $\pi \times 10^{-9}$ & 3 &$4.7\times10^{-13}$   & 1200 & 
weakly time-variable \\
11 & 0.6     &  $10^{-8    }$ & 3 &$6.0\times10^{-12}$   & 3500 & steady state \\
12 & 0.6     &  $\pi \times 10^{-8}$ & 3 &$4.7\times10^{-11}$   & 5000 & steady state\\
13 & 0.6     &  $  10^{-7}$   & 3 &$3.5\times10^{-10}$   & 7000 & steady state \\

         &                &   &            &      \\
14 & 0.6     &  $\pi \times 10^{-8}$ & 10 &$ 3.1\times10^{-10}$   & 7000 & steady state\\

         &                &   &            &      \\
15 & 0.4     &  $\pi \times 10^{-8}$ & 0 &$6.2\times10^{-15}$   &  3000 & complex wind\\
16 & 0.4     &  $  10^{-7}$   & 0 &$1.6\times10^{-13}$   &  6000 & complex wind\\

         &                &   &            &      \\
17 & 0.8     &  $\pi \times 10^{-8}$ & 0 &$6.2\times10^{-11}$   & 2500 & complex wind \\
18 & 0.8     &  $  10^{-7}$   & 0 &$6.3\times10^{-10}$   &  7500 & complex wind\\
19 & 0.8     &  $\pi \times 10^{-7}$ & 0 &$4.0\times10^{-09}$   &  14000  & complex wind\\ 

\hline
\end{tabular}

\end{center}
\normalsize
\end{table*}

Our simulations suggest that there are two kinds of flow that might
arise from luminous accretion disks (Proga, Drew, \& Stone 1997).
We describe a representative example
of each of these two types of outflow in some detail first (sections 4.1 and
4.2).  These are followed by a limited parameter survey in which we focus on 
the effects of varying three key parameters -- the disk luminosity,  
the relative luminosity of the CS ($x$) and the force multiplier index
$\alpha$.  Finally we draw attention to the role of some of the model 
assumptions in section 4.4.

\subsection{A complex weak wind case }

\begin{figure*}
\begin{picture}(180,590)
\put(0,0){\includegraphics{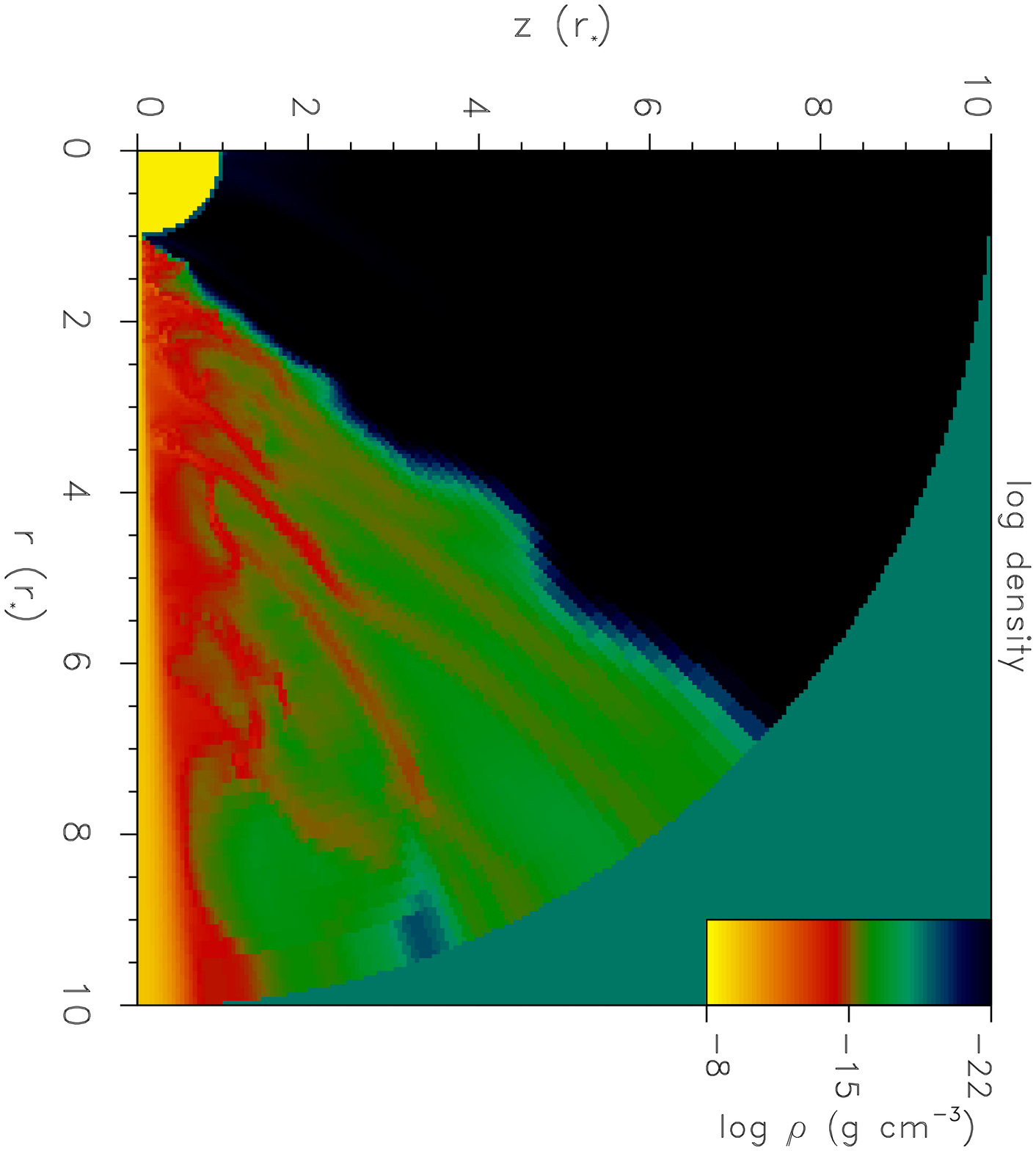}}
\put(0,205){\includegraphics{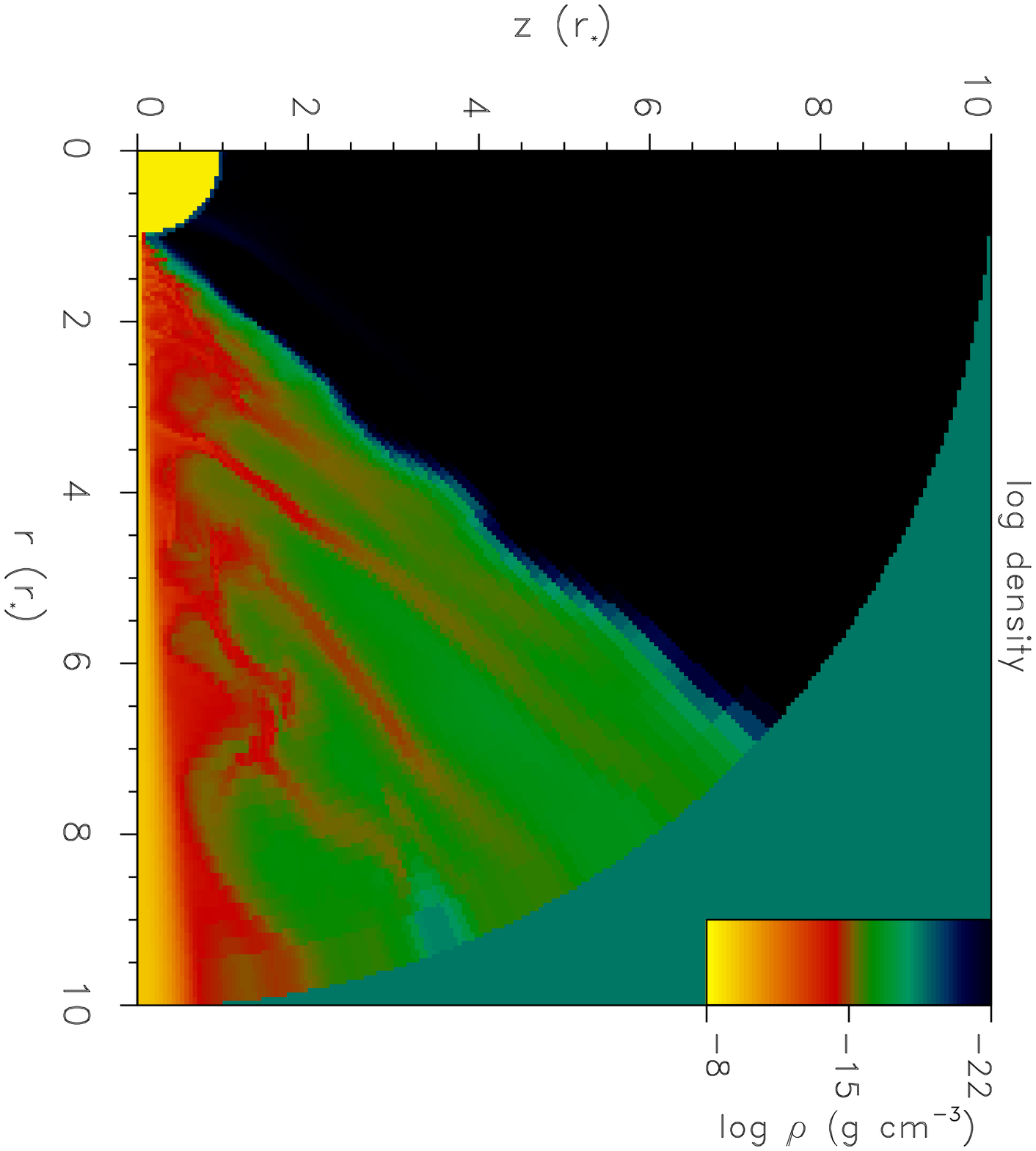}}
\put(0,410){\includegraphics{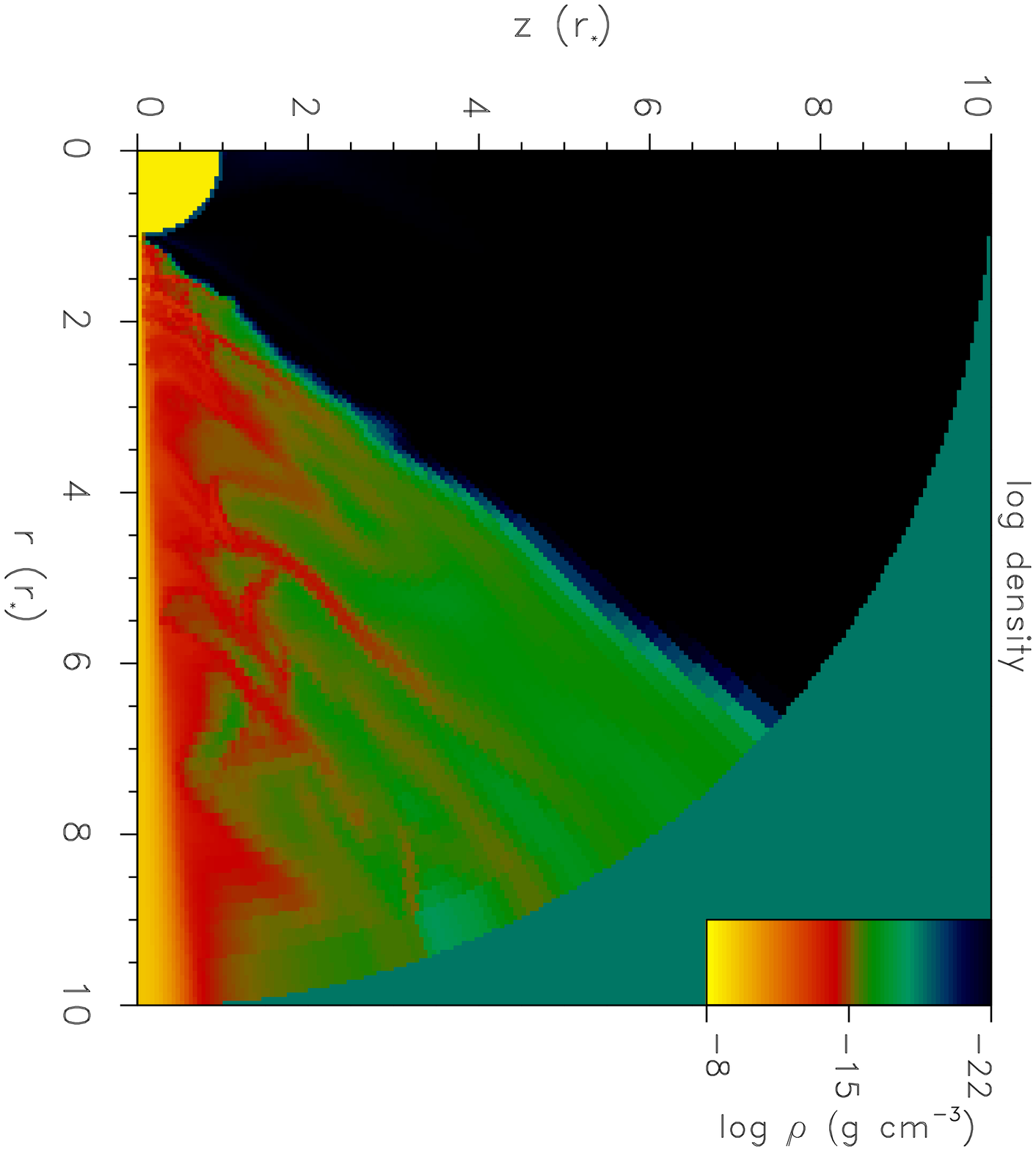}}
\put(90,0){\includegraphics{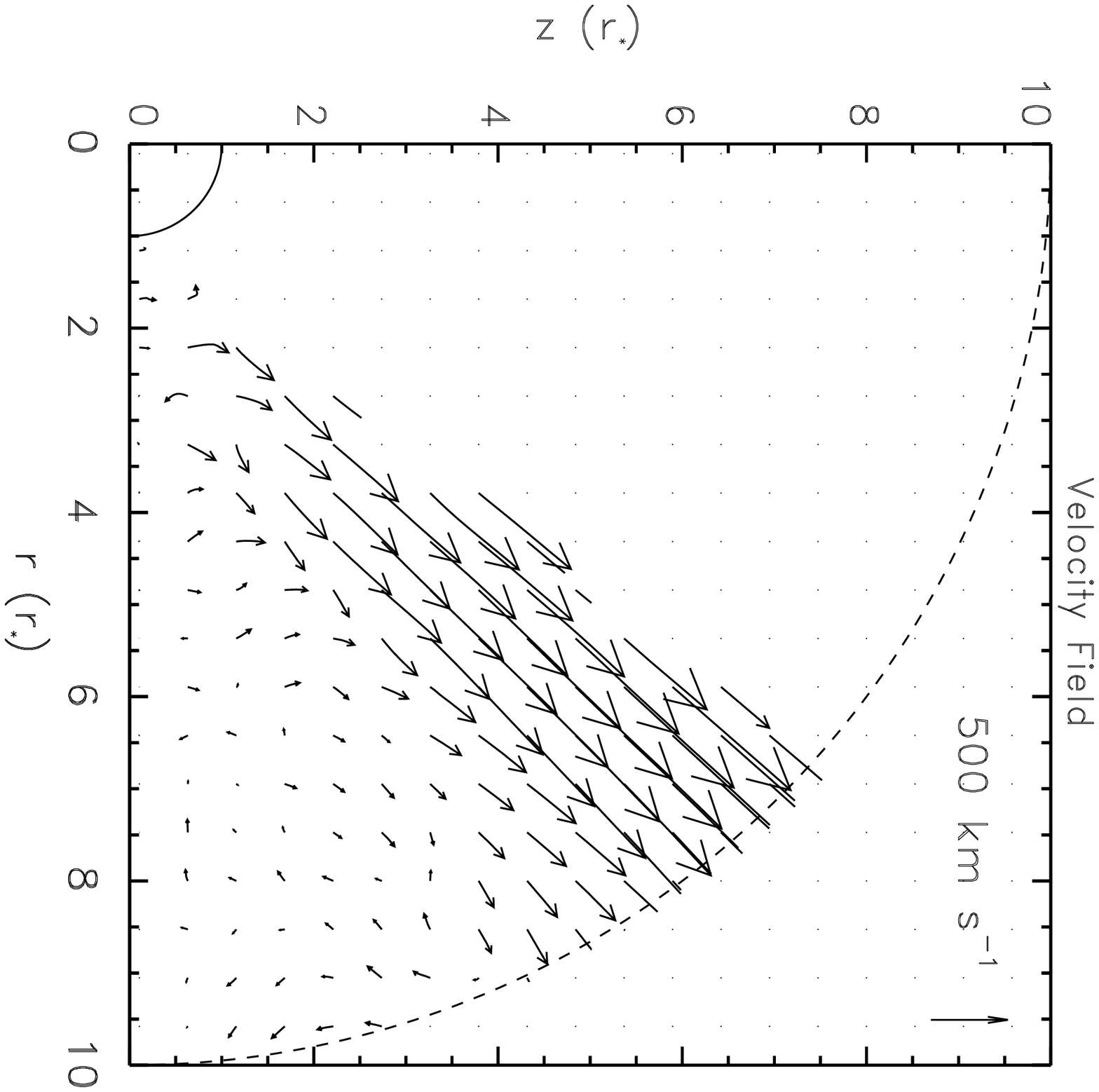}}
\put(90,205){\includegraphics{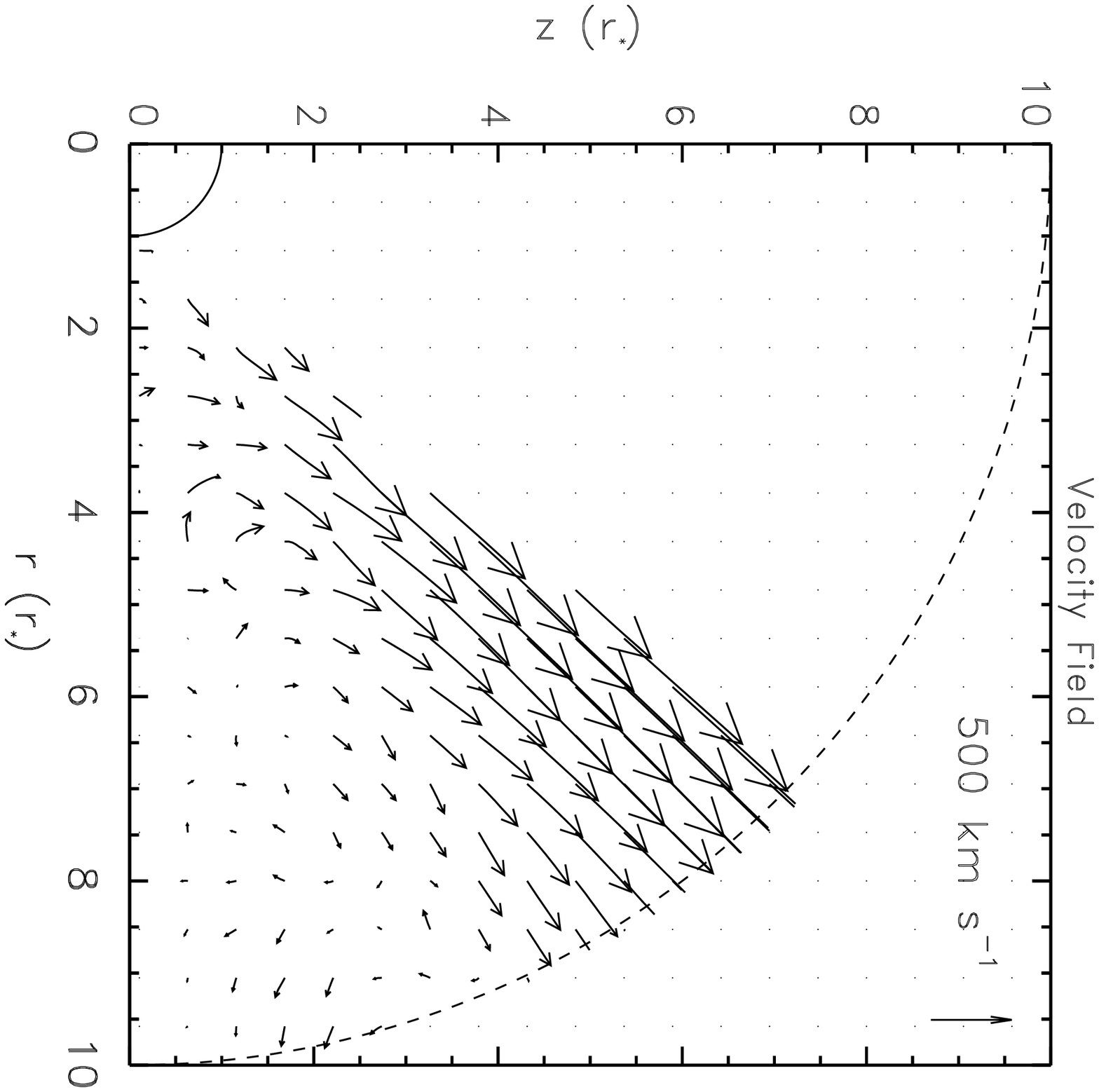}}
\put(90,410){\includegraphics{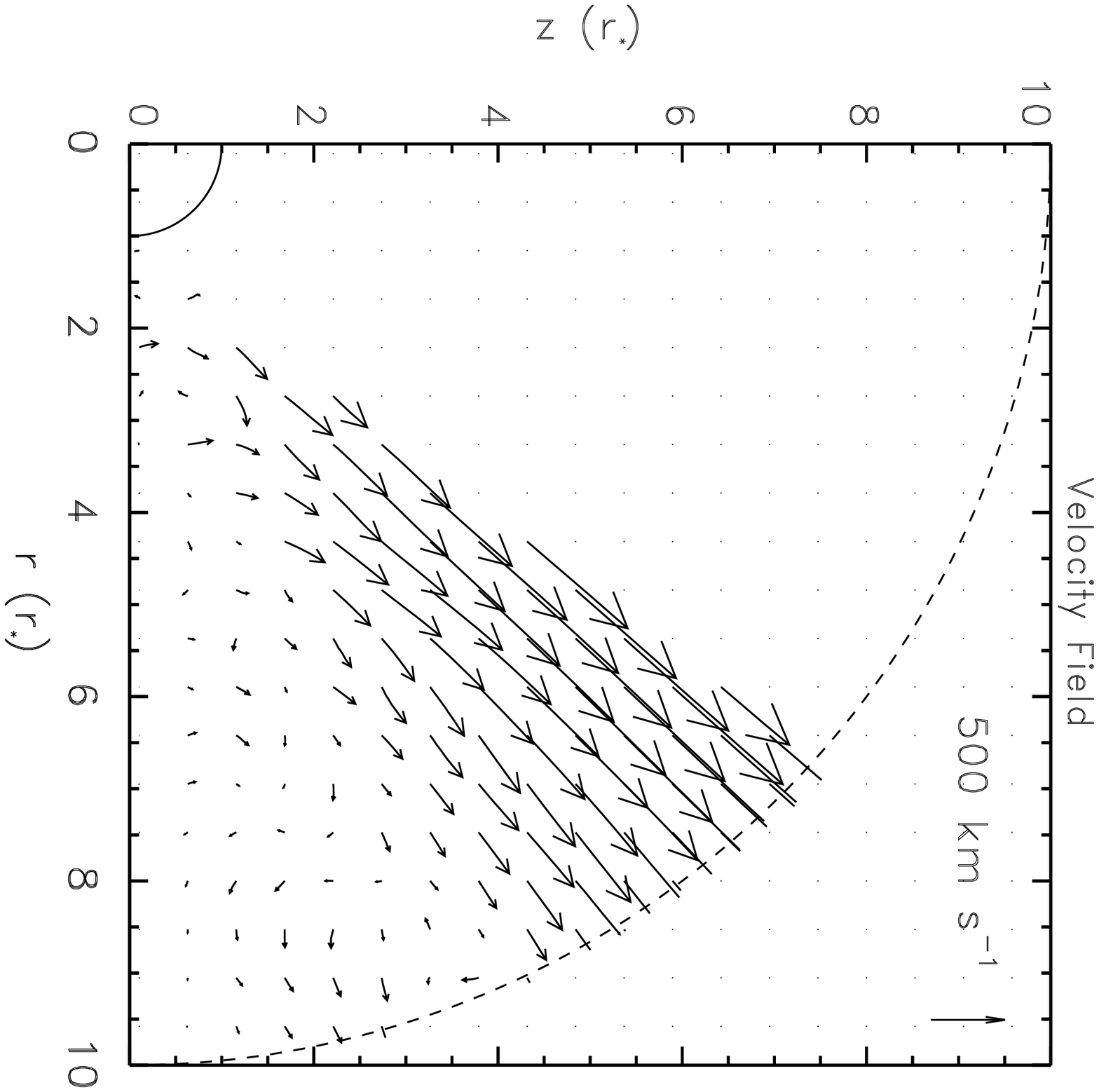}}
\end{picture}
\caption{A sequence of density maps (left) and velocity fields (right)
from run 2 after 208, 226 and 239 $\tau$ (upper, middle and lower panels).  
Run 2 is the example of unsteady flow discussed in detail in section 4.1.
Note the time-dependent fine structure in density and changing velocity
vectors at polar angles greater than $\sim65^o$. 
}
\end{figure*}  

We begin the presentation of our results by describing the properties
and behaviour of our model in which $\MDOT_a = 10^{-8}$~\MSUNYR and the
central star is assumed to be dark ($x = 0$).  This is our model 2
(see Table 2).

Figure~2 presents a sequence of density maps  and velocity fields
(left and right hand panels) from model 2 plotted in the $r,z$ plane.  
The figure displays the results of our high resolution ($200 \times 200$
grid points) run.  The length of the arrows in the right-hand panels is
proportional to $(v_r^2 + v_\theta^2)^{1/2}$.  The pattern of the
direction of the arrows is an indication of the shape of the
instantaneous streamlines.  After $\sim 10$ time units ( we define the
time unit as the orbital period at the surface of the CS, $\tau
=\sqrt\frac{r_\ast^3}{GM_\ast}~=~2.88~ \rm sec$) the disk material fills
the grid for $\theta \gtappeq 45^o$ and it remains in that region for the
rest of the run. Figure~2 shows that the high density region usually
corresponds to regions of low velocity.  The variation in the
orientation of the velocity arrows in the righthand panels indicates
the flow is time-dependent and, moreover, it is clear that in some
cases negative radial velocities (i.e. infall) are possible.  The time
dependence persists even after 80 $\tau$.  However it is important to
note as discussed below that the gross properties of the flow (such as
the mass loss rate), settle down to steady time-averages when averaged
over timescales on the order of 100 sec ($\sim 30\tau$) or more.

In model 2, the flow is complex with a few filaments sweeping outwards, 
typically, and various knots and clumps of gas moving both upwards and
downwards.  The direction and speed of motion at any one position is apt 
to change unpredictably with time;  the velocity magnitudes at $\theta >
60^o$ are typically less than $\sim100~\rm km~s^{-1}$. In the flow there is 
also a region where the material moves in a quite organized fashion. 
For $45^o \ltappeq \theta \ltappeq 60^o$ beyond $\sim3r_\ast$  
the material moves along nearly straight trajectories and leaves the outer 
boundary of the grid with velocities ranging from $\sim 300$ up to 
$\sim 1100~\rm km~s^{-1}$.  However, at no location in even this part of the   
flow does the velocity ever exceed the local escape velocity (which
decreases from $4280~\rm km s^{-1}$ at $r = r_\ast$ to $1350~\rm km s^{-1}$
at $r = 10r_{\ast}$).  This does not mean that the mass loss will necessarily
stall at a larger radius.  

To investigate the nature of the outflow on larger scales, we have
calculated this model on a computational domain that is ten times
larger in the radial direction, i.e. it extends from $r_\ast$ to
$100r_\ast$.  We use a grid of 100 angular zones and 150 radial zones
for this model, so that the numerical resolution in the inner region of
the grid (i.e. $r \leq 10r_\ast$) is identical to our standard case.
The results of this calculation show that the integrated mass loss rate
at $100r_\ast$ is the same as that at $10r_\ast$.  Moreover, the fast
stream does continue to accelerate beyond $10r_{\ast}$, so that $v_r$
rises from $\sim$1100 km s$^{-1}$ at $10r_{\ast}$ up to 1300 km
s$^{-1}$ or so at $100r_{\ast}$ (which is well above the escape
velocity at this point).  That the fast outflow eventually exceeds
escape velocity is not suprising since at large radii the density of
the flow is so low that the radiation force reaches its maximum value
set by $M_{max}$: thus both gravity and the radiation force will scale
with radius as $r^{-2}$.  Consequently, not only can the total force
never change sign (because the ratio of the radiation force to gravity
is constant), but also the total force decreases in magnitude rapidly,
so that the flow velocity no longer changes significantly.  More
importantly, the model on the larger grid indicates the mass loss is
completely dominated by material arising from the inner ($r <
10r_{\ast}$) region of the disk.

\begin{figure}
\begin{picture}(100,480)
\put(0,0){\includegraphics{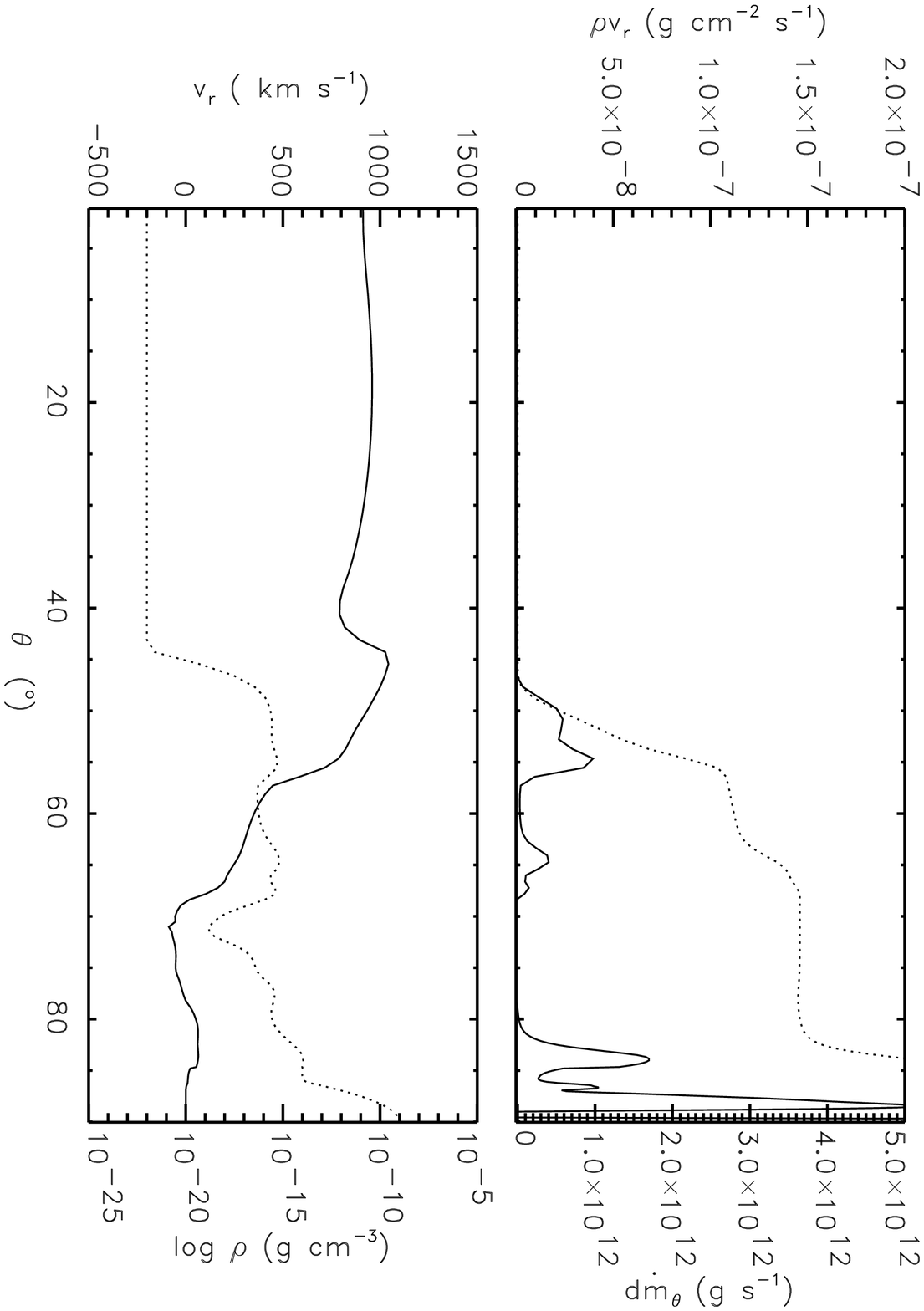}}
\put(0,160){\includegraphics{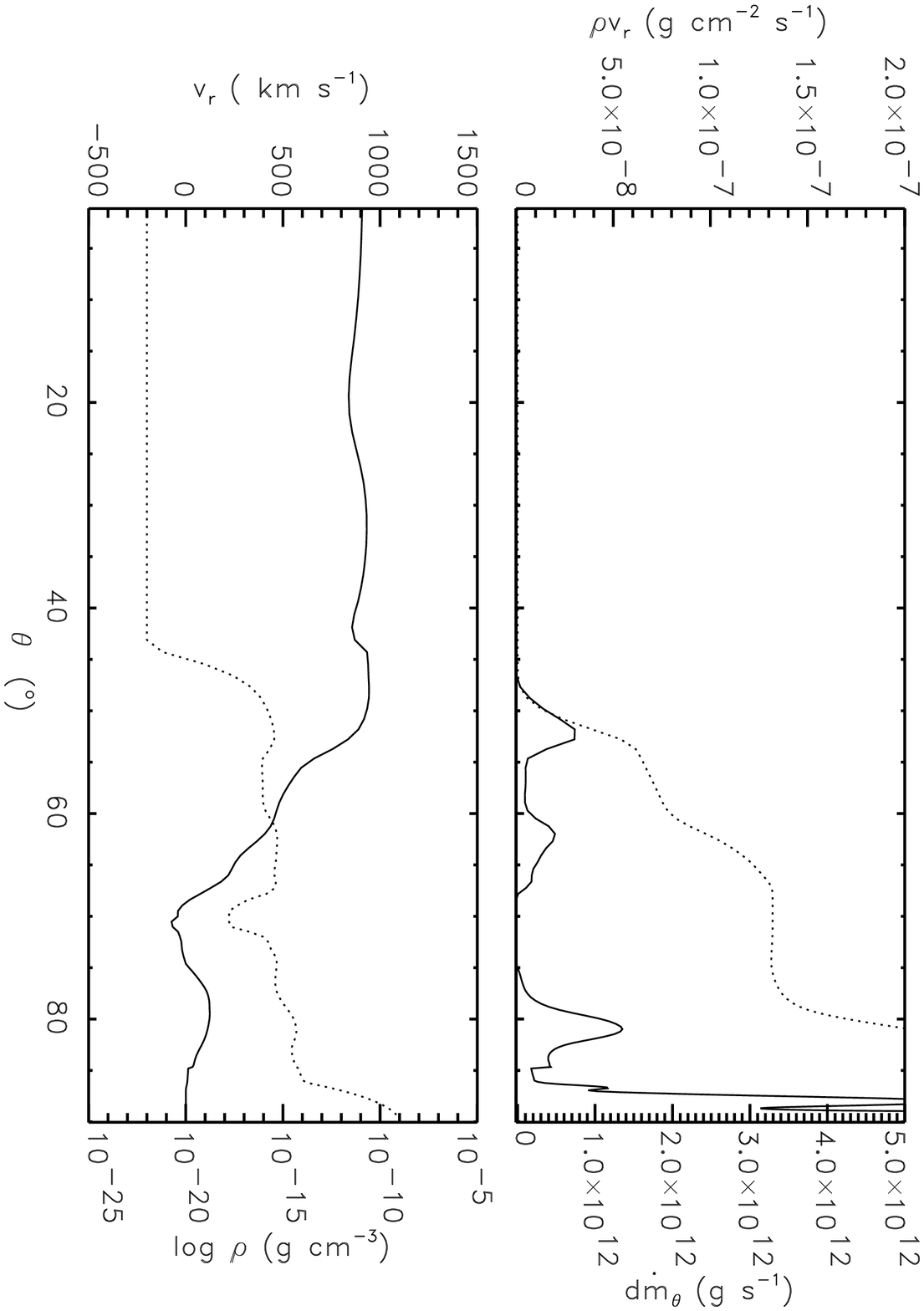}}
\put(0,320){\includegraphics{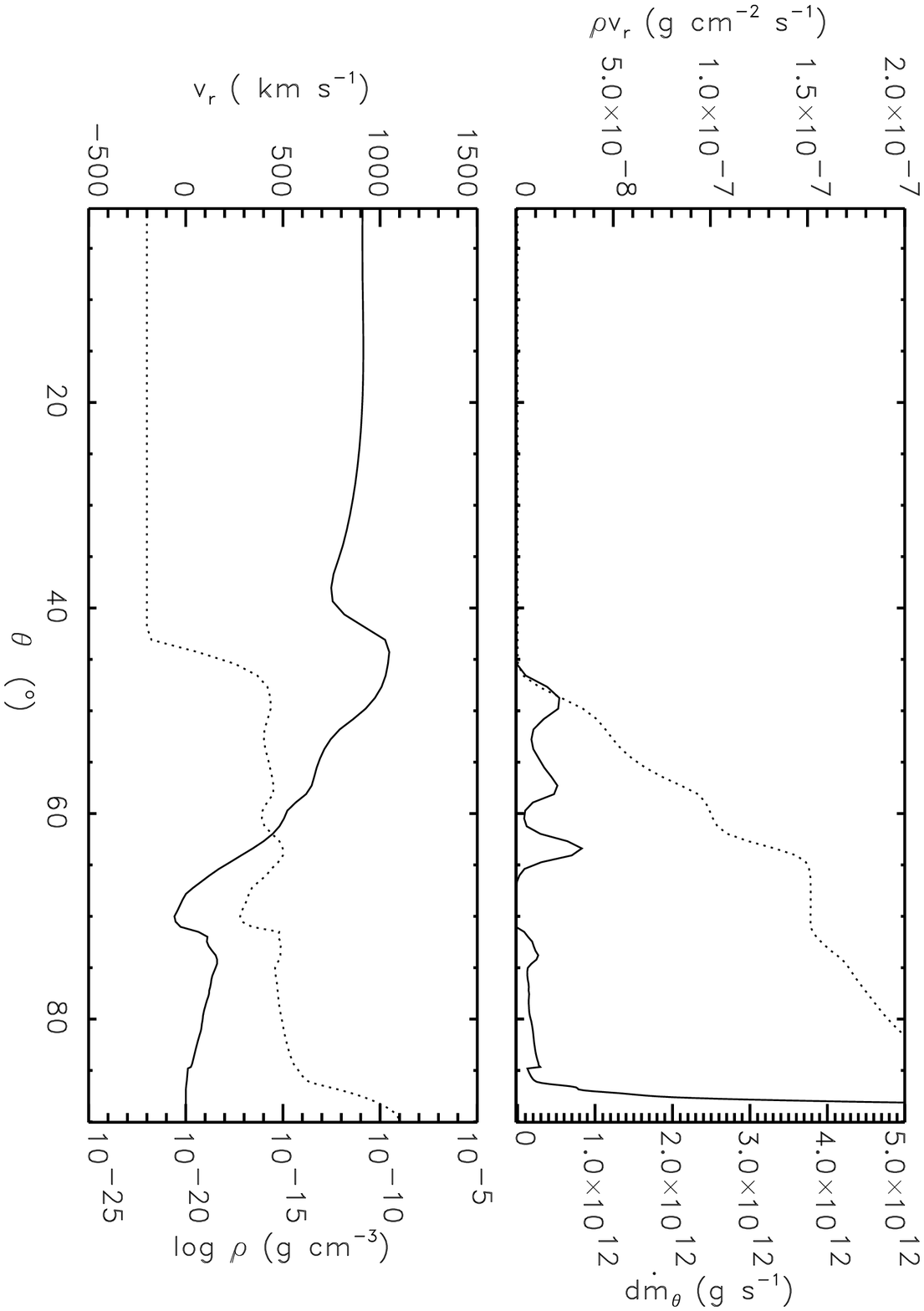}}
\end{picture}
\caption{
A sequence showing how quantities at the outer boundary in
model 2 change as a function of time.  As in Figure~2 the top, middle
and lower panel pairs refer to the times 208, 226 and 239 $\tau$.  In each
panel the ordinate on the left hand side describes the solid line, while
the ordinate on the right hand side refers to the dotted line.  All curves
show considerable structure that is time-variable.  The cumulative
function $d\dot{m}(\theta)$ (dotted line, upper panel in each pair) is 
smoother and less variable, illustrative of the manner in which global 
quantities tend toward steady time averages.  
}
\end{figure}  

Next we consider the time and angle dependence of the gross properties of the 
flow at large radii.  Figure~3 is a plot of the angular dependence of 
density, radial velocity, mass flux density, and accumulated mass loss rate 
at $r = r_o = 10 r_{\ast}$ at the same times as Figure~2.
The accumulated mass loss rate is given by:
\begin{equation}
d\dot{m}(\theta) =
4 \pi r_o^2 \int_{0^o}^\theta \rho v_r \sin \theta d\theta.
\end{equation}
The gas density is a very strong function of angle for $\theta$ between
$\sim 90^o$ and $45^o$.  Between the disk mid-plane at $\theta = 90^o$ and 
$\theta \sim
85^o$, $\rho$ drops by $\gtappeq 6$ orders of magnitude, as might be expected
of a density profile determined by hydrostatic equilibrium (see equation 11). 
For $45^o \ltappeq \theta \ltappeq 85^o$, the wind domain, $\rho$ varies 
between $10^{-17}$ and $10^{-15}~\rm g~cm^{-3}$.  For $\theta \ltappeq 45^o$, 
density again decreases exponentially, but this time to so low a value that 
it becomes necessary to replace it by the numerical lower limit $\rho_{min}$.
The region with $\rho \le \rho_{min}$ is not relevant to our analysis as it 
has no effect on the disk flow.  The radial velocity at $10r_\ast$ varies 
around zero with an amplitude $\ltappeq 100~\rm km~s^{-1}$ for $65^o 
\ltappeq \theta \ltappeq 90^o$.  Over the angular range  $65^o > \theta >
45^o$, $v_r$ increases from $\leq 100$ up to $1200~\rm km~s^{-1}$.

The cumulative mass loss rate is negligible for $\theta \ltappeq 45^o$
because of the very low prevailing gas density.  Beginning at $\theta 
\gtappeq 45^o$, $d\dot{m}$ increases to $\sim 4\times 10^{12}~\rm g~s^{-1}$ 
at $\theta \approx 83^o$.  Then, in the region close to 
the disk plane, where the gas density starts to rise very sharply and where 
the motion is typically more complex, the cumulative mass loss rate is 
subject to enormous fluctuations (some of which may even be negative!).  In 
the example shown as figure 3, the total mass loss rate through the outer 
boundary, $\MDOT_{tot}=$ $d\dot{m}(90^o)$ reaches $\sim 5 \times 10^{13}~\rm 
g~s^{-1}$.  This figure is most certainly dominated by the contribution from 
the slow-moving region very close to the disk mid-plane -- a contribution 
that is very markedly time-dependent.

\begin{figure}
\begin{picture}(100,170)
\put(0,0){\includegraphics{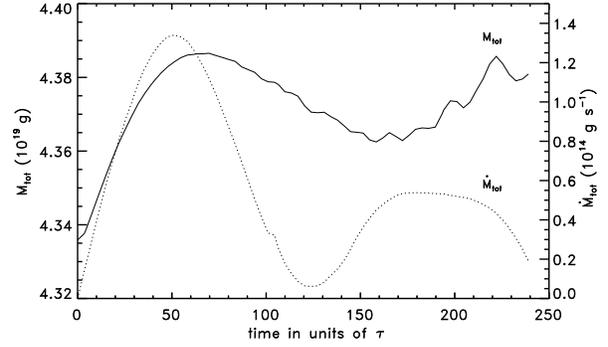}}
\end{picture}
\caption{
The time evolution of the total mass on the grid and the
total mass loss rate ($d\dot{m}(90^o)$) for model 2.  The strong modulation 
of the total mass loss rate is dominated by the slow and highly variable flow 
component close to the disk plane (see figures 2 and 3).  In this case,
the relatively steady fast stream at $\theta \sim 65^o$ contributes only
$\sim$10\% of the total mass loss. 
}
\end{figure}

To provide some insight into the time dependence, figure~4 shows the time 
evolution of volume-averaged quantities for run 2. The total mass on the 
grid, $M_{tot}$, is subject only to small changes.  It increases by 
$\sim 1 \%$ during the first 60 $\tau$, and then decreases again, dropping 
back to 1.006 of the initial value, $M_{tot}(0)$, by 170 $\tau$.  After this
time $M_{tot}$ starts to fluctuate between $\sim$1.006 $M_{tot}(0)$ and 
$\sim 1.01 M_{tot}(0)$.  By contrast, the total mass loss rate is seemingly a 
strong function of time.  Initially, it rises steeply, peaking at 
$1.3\times 10^{14}~\rm g~s^{-1} $ at 50 $\tau$. Then it plummets to 
$5\times 10^{12}~\rm g~s^{-1} $ at $\sim120\tau$ and starts oscillating with 
decreasing amplitude.  These large swings are entirely a consequence of the 
complex character of the flow close to the disk mid-plane.  A much steadier,
and consistently positive, cumulative mass loss rate is achieved if the 
integration over polar angle is stopped at $\theta \ltappeq 70^o$
(see figure~3).  A further justification for stopping the integration at this 
angle, is that $d\dot{m}$ is then the mass loss rate associated with just the 
hypersonic outflow that easily escapes the system.

\subsection{ A strong steady outflow model }

We find that unsteady outflow such as that described above in section 4.1
persists as long as the disk radiation field is dominant (small $x$ in
our parameterisation).  However, as the radial component of the radiation 
field is increased with respect to the latitudinal ($\theta$) component, by
adding in light from a central star (CS), we find that the volume occupied by 
unsteady outflow diminishes.  Indeed, in the presence of a strong radiation 
force due to the CS, a disk wind can even settle into a steady state.  

\begin{figure*}
\begin{picture}(180,220)
\put(0,0){\includegraphics{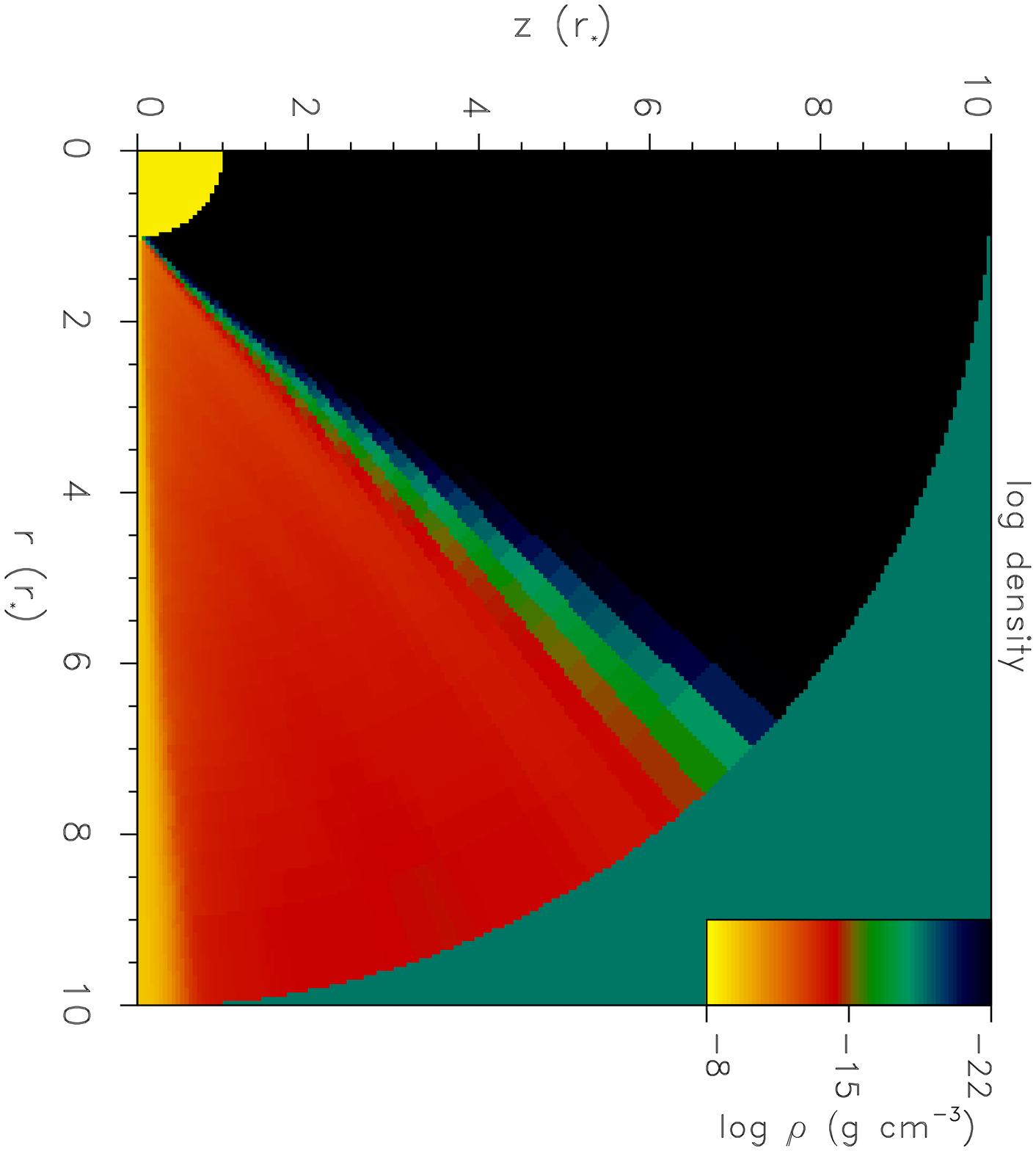}}
\put(90,0){\includegraphics{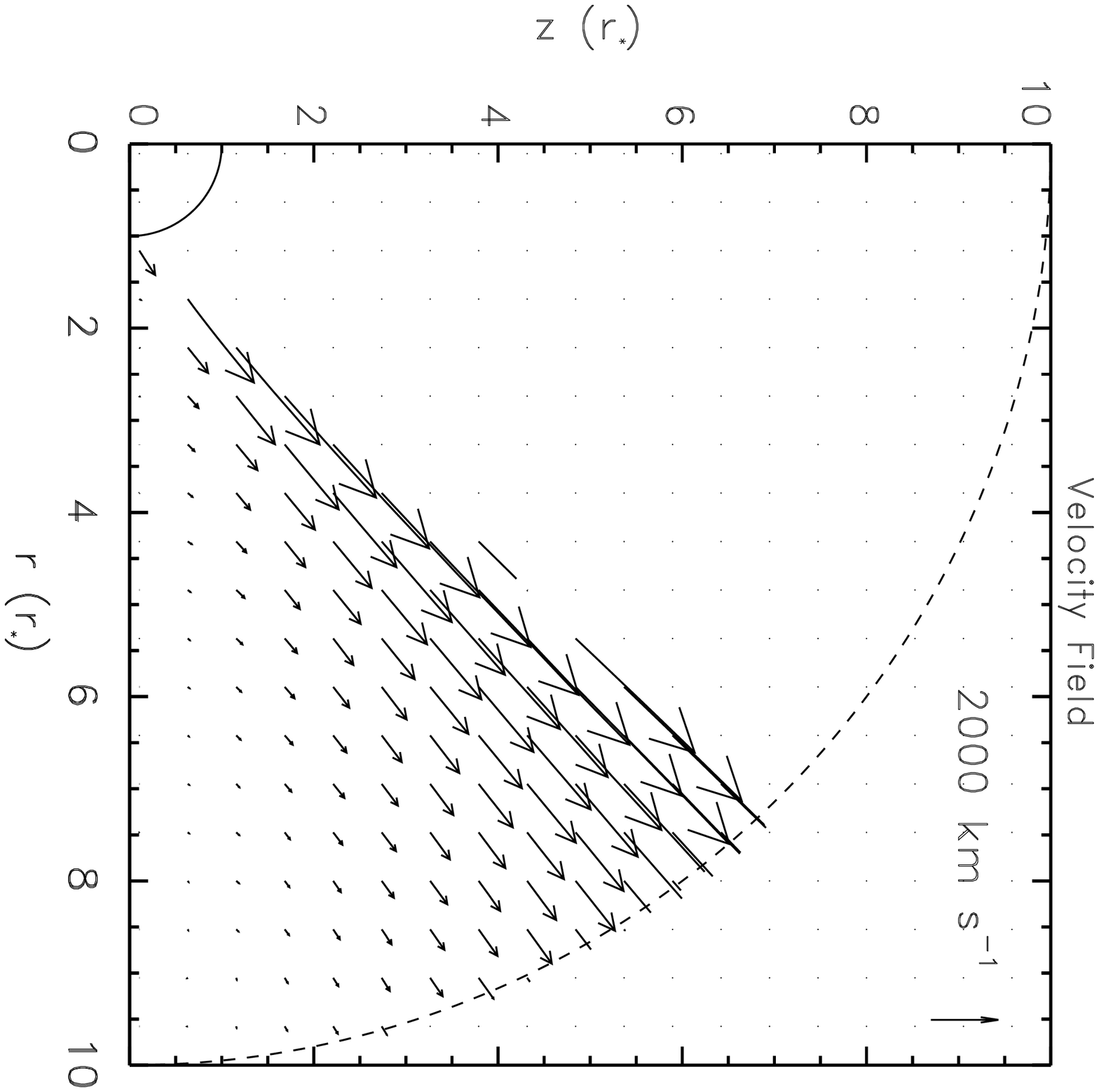}}
\end{picture}
\caption{
The density map and velocity field for model 8 after 240 
$\tau$.  Note the absence of fine structure in the density map and the
regularity of the velocity field. 
}
\end{figure*}  

\begin{figure}
\begin{picture}(100,170)
\put(0,0){\includegraphics{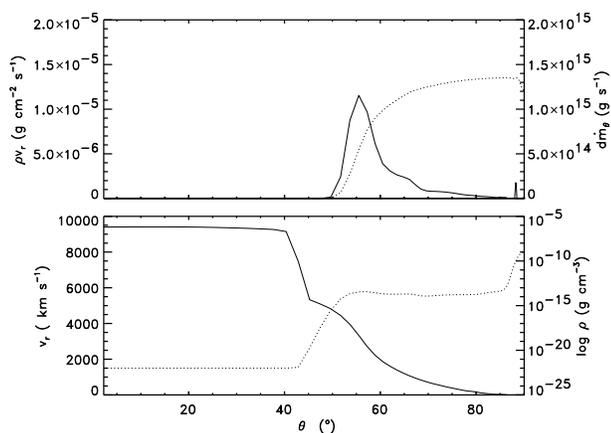}}
\end{picture}
\caption{
Quantities at the outer boundary in model 8 after 240~$\tau$.
The ordinate on the left hand side of each panel refers to the solid line, 
while the ordinate on the right hand side refers to the dotted line.
Of particular note is the pronounced $\rho v_r$ peak at $\theta \sim 55^o$.
This is associated with the fast stream that contributes $\sim$90\%
or so of the total mass loss.
}
\end{figure}  

\begin{figure}
\begin{picture}(100,170)
\put(0,0){\includegraphics{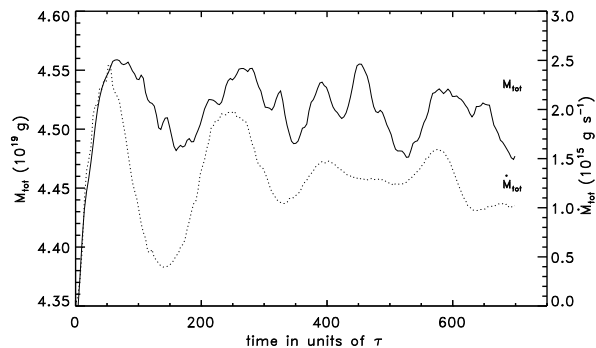}}
\end{picture}
\caption{
The time evolution of the volume-averaged quantities for 
model 8 (to be compared with model 2 shown in figure 4).  In this case the 
total mass loss rate is much more settled, reflecting a very nearly steady 
flow and the dominant role of the fast stream at $\theta \sim 55^o$.
}
\end{figure}  

Model 8 is a contrasting example of a strong outflow in a steady state. 
In it, we set the CS luminosity equal to the intrinsic disk luminosity 
(i.e. $x = 1$), and chose $\MDOT_a = \pi \times 10^{-8}$~\MSUNYR.   The 
remaining model parameters and the initial conditions are as specified in 
Tables~1 and 2.
Figure~5 is the density map and velocity field for this model after 
240$\tau$.  The flow is almost in a steady state by then, with the gas 
density a smooth function of position.  The flow may be described as 
organized and regular.  Small changes with time still occur, but only very 
close to the disk plane. 

Figure~6 presents the wind properties as function of $\theta$ at $r=10 r_\ast$
after 240$\tau$.  On this surface, the flow density varies between $10^{-14}$ 
and $10^{-13}~\rm g~cm^{-3}$ for $50^o \le \theta \le 85^o$. 
Within the same $\theta$ range, $v_r$ increases from low values on the order
of 100~$\rm km~s^{-1}$, up to $\sim 4000~\rm km~s^{-1}$ as $\theta$ decreases.
The accumulated mass loss rate is $1.3\times10^{15}~\rm g~s^{-1}$ at
$\theta \sim 80^o$.  This is a factor of $\sim$400 increase with respect
to the mass loss rate obtained in model 2, for just a factor of $2\pi$ 
increase in total luminosity.

Figure~7 shows the time evolution of volume-averaged quantities in model 8.
This should be contrasted with the equivalent figure for model 2 (figure 4). 
As in model 2 all the quantities plotted are subject to fluctuations of the 
wind properties near the disk plane, where the flow does not quite settle 
into a steady state.  However the magnitude of these fluctuations of the
total mass loss rate has collapsed from a factor of 10 (model 2, figure 4) to 
around 1.5 (model 8, figure 7).

\subsection{ Parameter survey }

As is fitting for a first exploration of radiation-driven wind models from 
disks, we aim to examine only the parameter space of our models that will 
define the major trends in disk wind behaviour.  Table~2 lists the models 
considered.
We emphasise a survey of how the mass loss rate, outflow velocity and geometry
change with disk luminosity and relative CS luminosity.  In view of the
important formal role that the force multiplier index $\alpha$ is known to 
play in determining one-dimensional stellar wind solutions (i.e. $\MDOT_w 
\propto L^{1/\alpha}$), we have also calculated a few models in which the 
index $\alpha$ been set equal to the relatively extreme values of 0.4 and 0.8 
(see Gayley 1995).  So that we might focus on this dependence, we arbitrarily
hold $k$ and $M_{max}$ constant.

\begin{figure}
\begin{picture}(100,310)
\put(0,0){\includegraphics{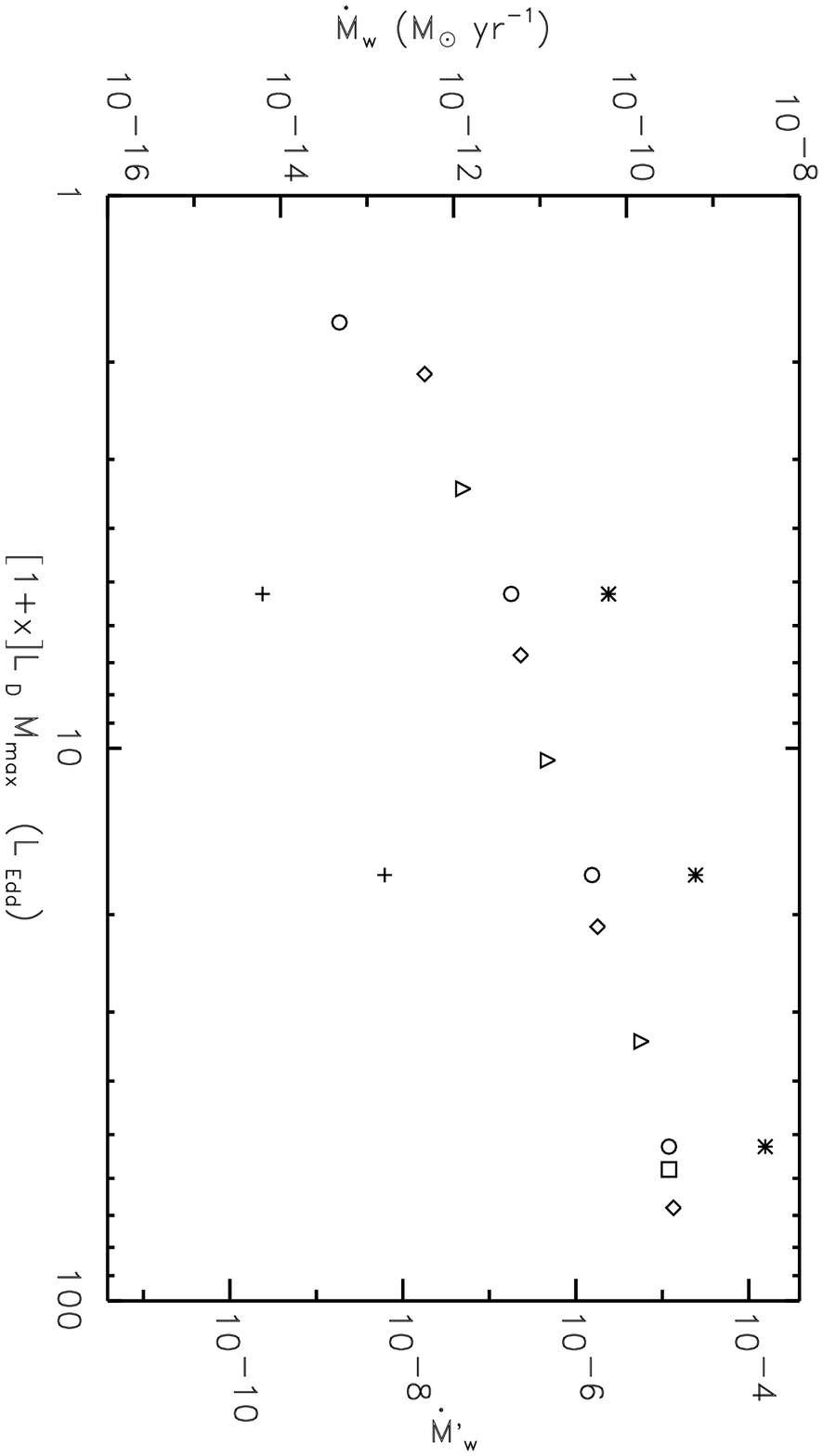}}
\put(0,155){\includegraphics{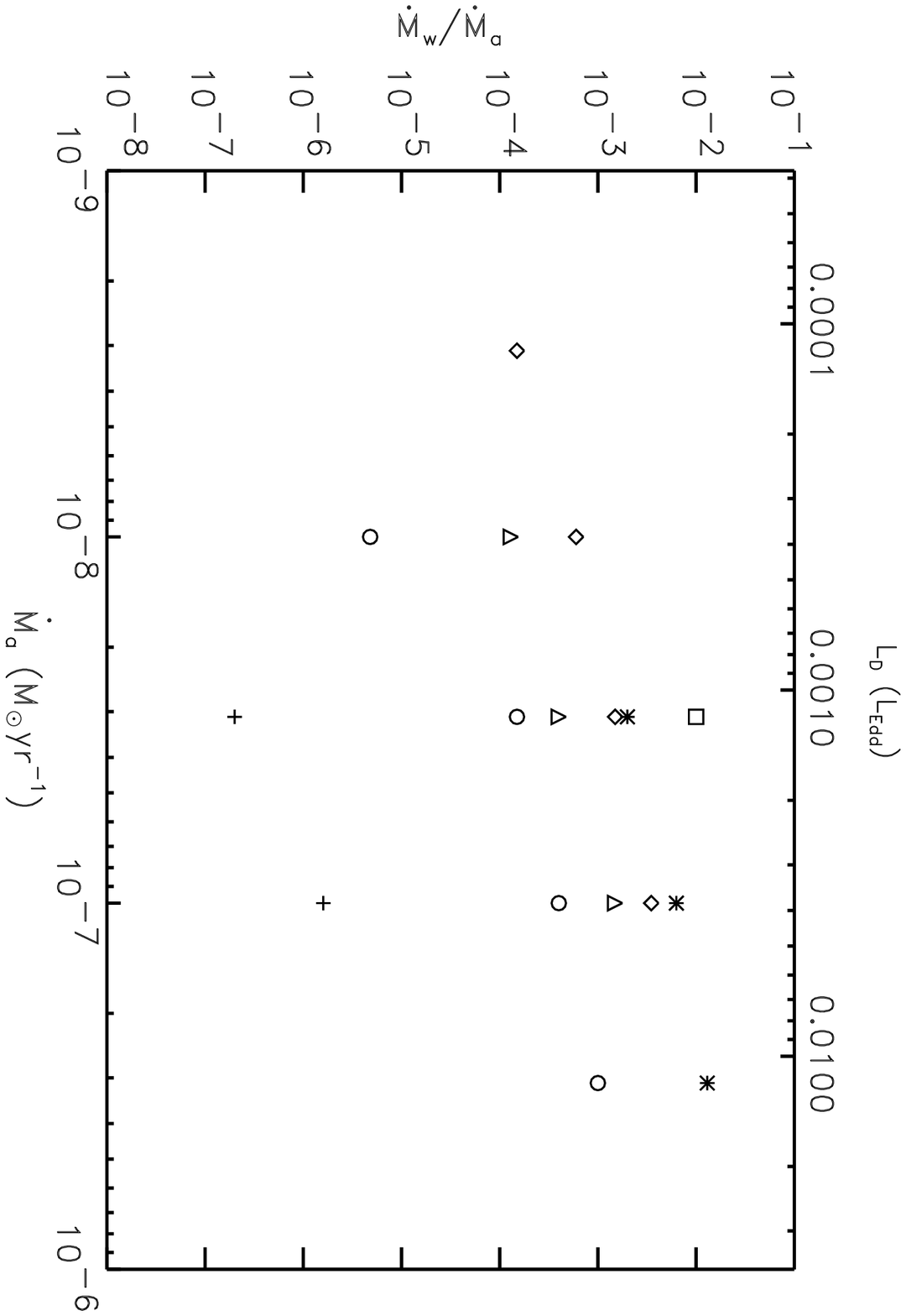}}
\end{picture}
\caption{
Model mass loss rates as functions of the mass accretion
rate (top panel) and effective total luminosity in units of the
Eddington luminosity (bottom panel).  All open symbols are for models with 
$\alpha=0.6$, the different shapes corresponding to different $x$ 
($x=0$ circles, $x=$1 triangles, $x=3$ diamonds, $x=10$, squares).
The crosses represent models for $\alpha=0.4$ and $x=0$ while
stars represent models for $\alpha=0.8$ and $x=0$.  Tables~1 and 2 
specify all other model parameters.  The alternative ordinate on the
right hand side of the lower panel is the dimensionless wind mass loss
rate parameter $\dot{M}_{w}'$ defined in equation 22, section 5.3.
}
\end{figure}  

\begin{figure}
\begin{picture}(100,310)
\put(0,0){\includegraphics{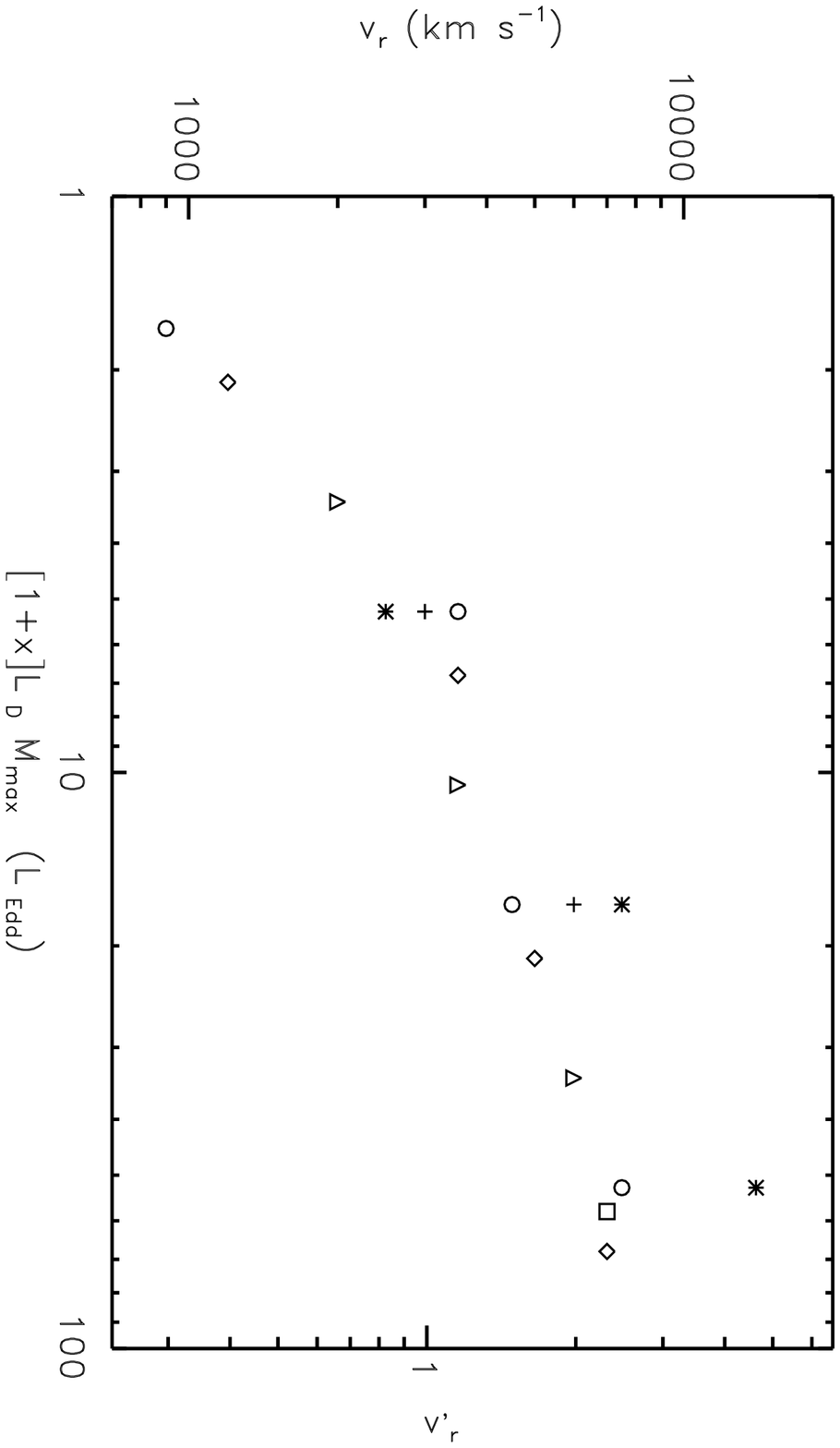}}
\put(0,155){\includegraphics{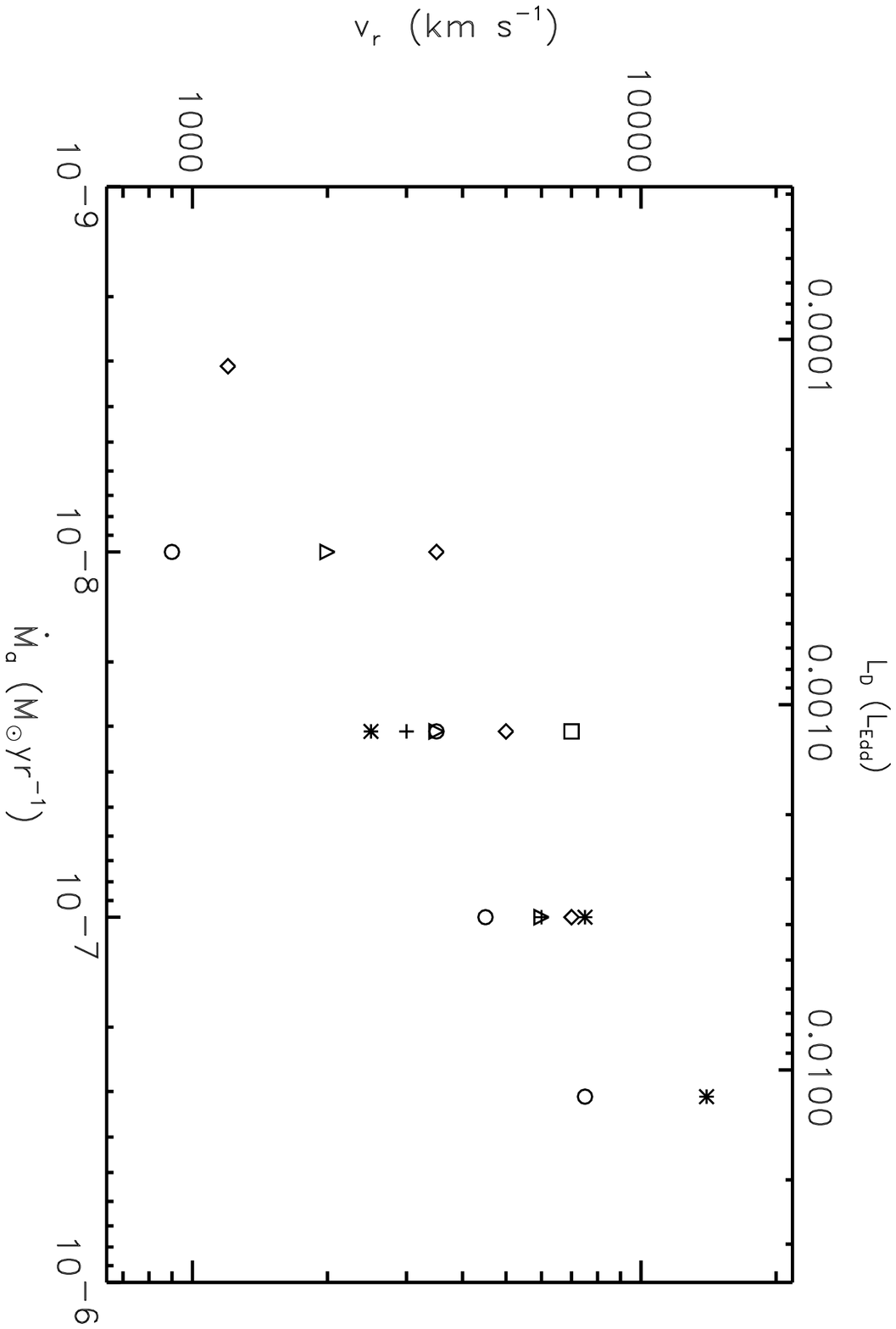}}
\end{picture}
\caption{
Typical fast stream velocities at the outer boundary
($r = 10r_{\ast}$) for all models as functions of the mass accretion
rate (top panel) and effective total luminosity in units of the Eddington
luminosity.  All symbols have the same meaning as in figure~8.  The
alternative ordinate on the right hand side of the lower panel is the
dimensionless velocity parameter $v'$ defined via equation 17 in section
5.3.
}
\end{figure}

In Figure~8 we show (a) the derived ratio, $\MDOT_w/\MDOT_a$, as a 
function of $\MDOT_a$ and (b) the wind mass loss rate, $\MDOT_w$, as a 
function of the total effective luminosity, $[1 + x] L_D M_{max}$,  for  
various $x$ and $\alpha$.  We define $\MDOT_w$ as the cumulative mass loss 
rate for the region well above the disk plane in which the highly supersonic, 
organized flow is located (i.e. the angular integral is stopped early enough 
to avoid the exponential density profile of the disk and any
lower-velocity complex flow component at $\theta$ near $90^o$).  The
total effective luminosity is the total luminosity of the system,  
$[1 + x] L_D$, multiplied by the maximum value of the force multiplier
$M_{max}$ (see the discussion leading to equation 4); it is measured in
Figure~8 in units of the classical Eddington value.  In Figure~8a it
can be seen that  $\MDOT_w/\MDOT_a$ is a very strong function of
$\MDOT_a$ for $x=0$.  At low $\MDOT_a$ there is virtually no disk wind
at all. The outflow turns on sharply for $\MDOT_w
\gtappeq~10^{-8}$~\MSUNYR, and then there is a flattening out of
$\MDOT_w/\MDOT_a$ to follow a power law of index $\sim 1.5$ for $\MDOT
\gtappeq~\pi \times 10^{-8}$~\MSUNYR.  A similar trend is apparent from
the $x=1$ and $x=3$ models, with the difference that increased $x$ at a
fixed mass accretion rate $\MDOT_a$ results in a higher mass loss
rate.  But when, instead, the absolute values of the calculated mass
loss rates are considered just as a function of the total effective
luminosity, this family of curves for different $x$ collapses, most
impressively, into a single curve.  This is shown in Figure 8b.  That
this occurs shows that $\MDOT_w$ is not itself sensitive to the
geometry of the driving radiation field (provided that $L_D M_{max}$ is
higher than the Eddington limit).  A further point to note from Figure
8b is that a disk together with CS will produce a fast wind for $\alpha
= 0.6$ if the effective luminosity is higher than $\sim$2 times the
Eddington limit.

\begin{figure*}
\begin{picture}(180,430)
\put(0,0){\includegraphics{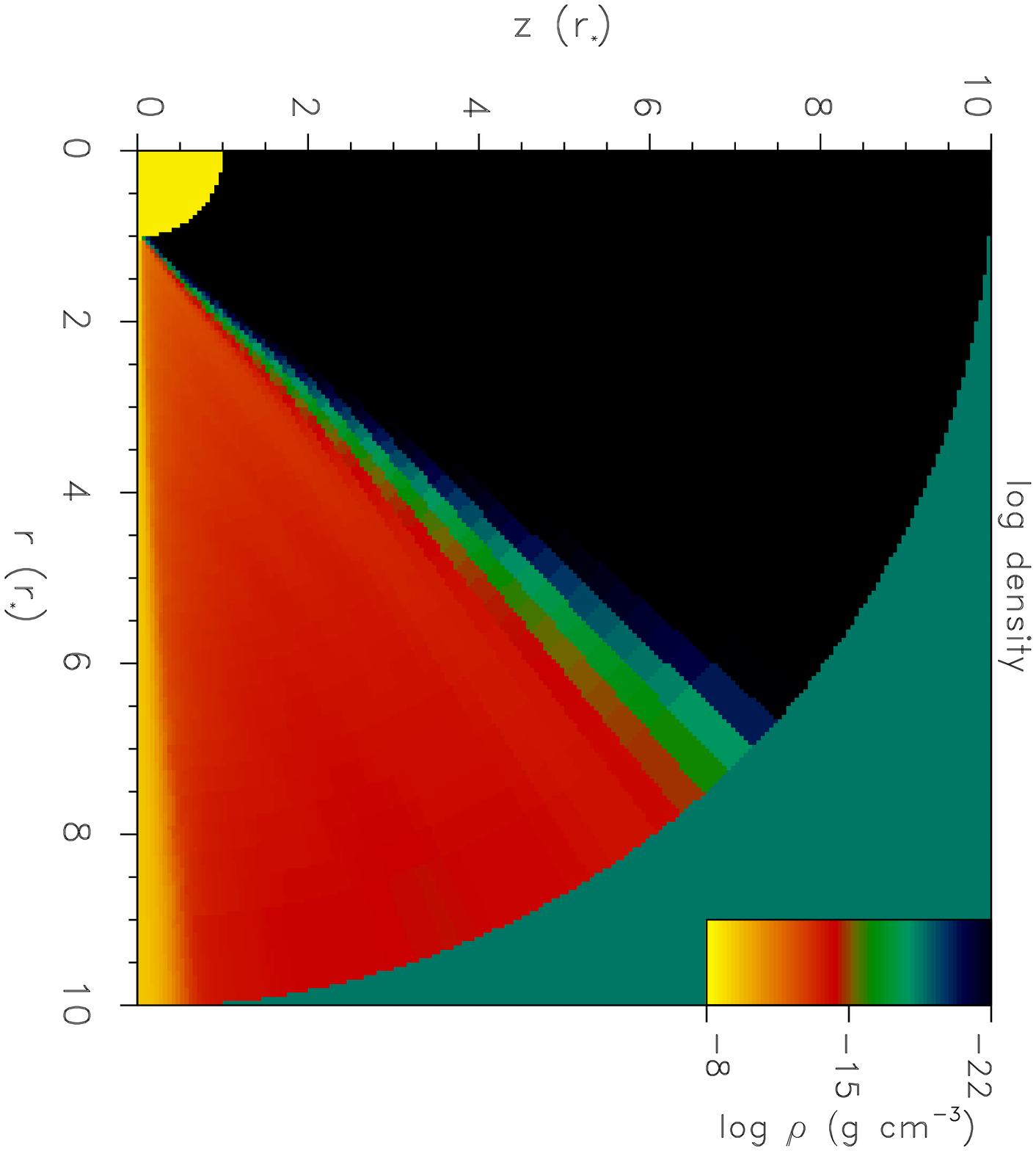}}
\put(0,210){\includegraphics{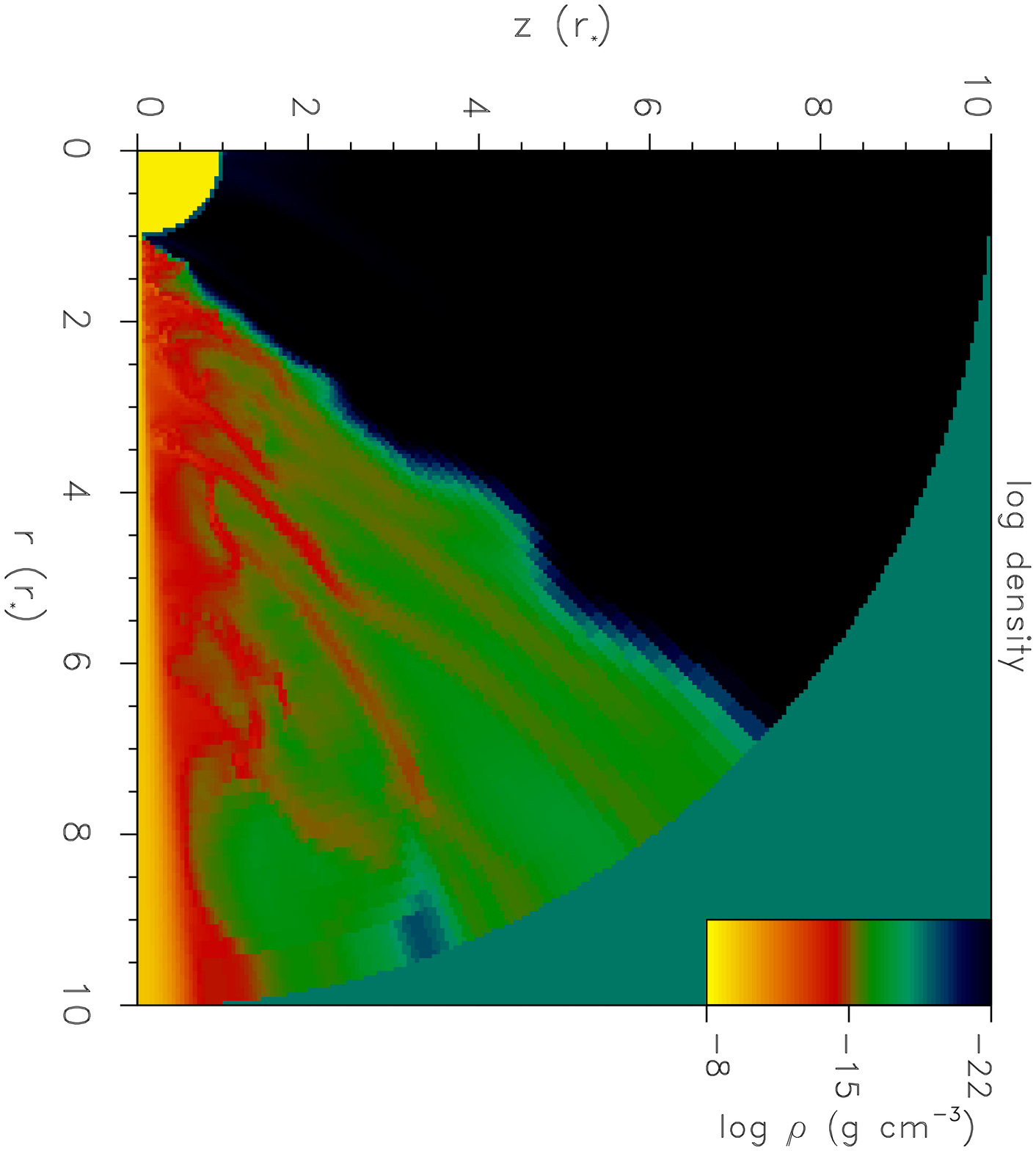}}
\put(90,0){\includegraphics{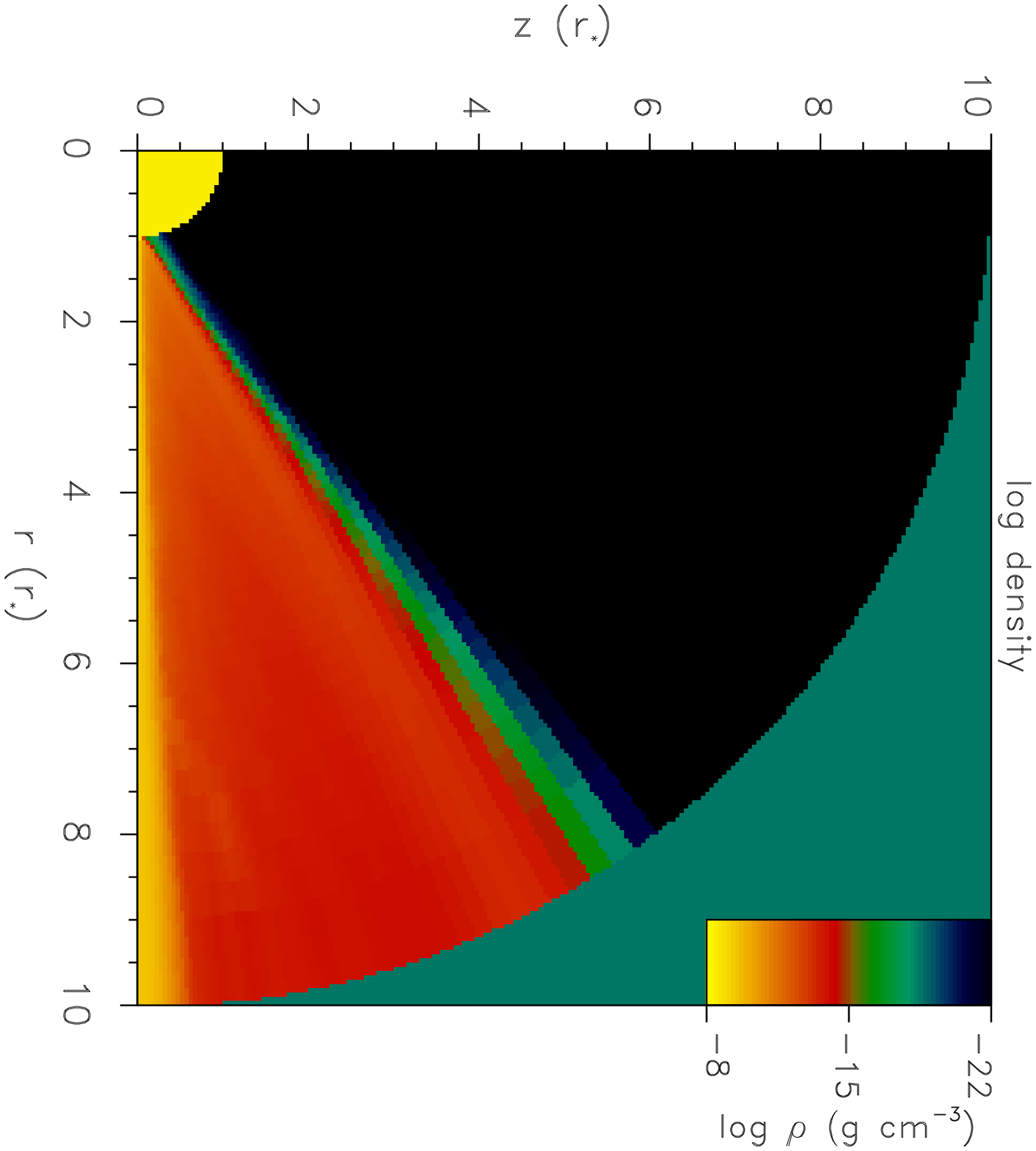}}
\put(90,210){\includegraphics{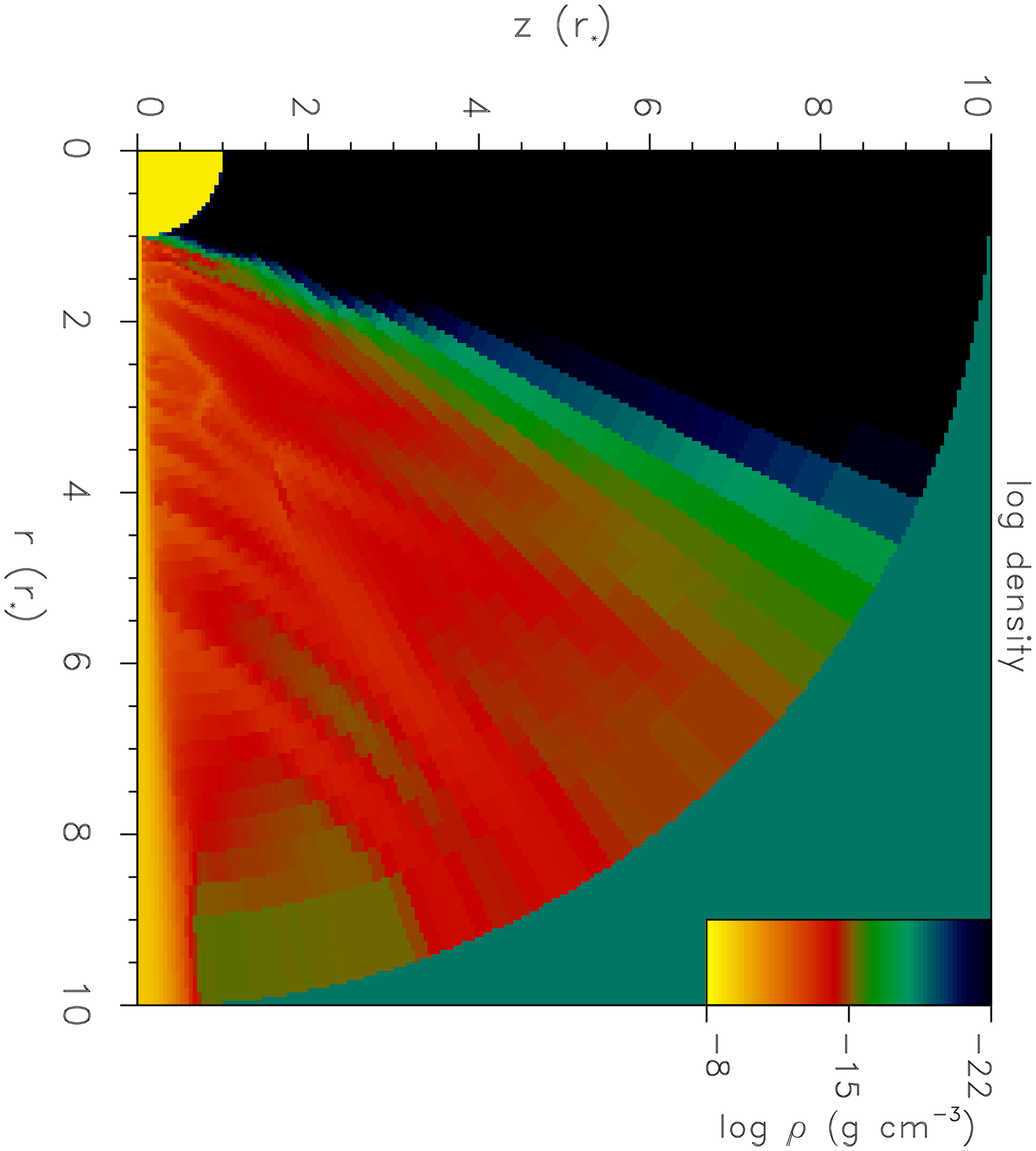}}
\end{picture}
\caption{
The top two panels, a and b, are density maps for models with
$x=0$ and $\MDOT_{a} = 10^{-8}$~\MSUNYR (model 2) and $\MDOT_{a} = \pi
\times 10^{-8}$~\MSUNYR (model 3).  The bottom two panels, c and d, are 
density maps for models with $\MDOT_{a} = \pi \times 10^{-8}$~\MSUNYR and
$x=1$ (model 8) and $x=3$ (model 12).  The top two panels show the
effect on the outflow geometry of increasing the disk luminosity alone,
while the top right and bottom two panels show the effect of adding in an 
increasingly larger stellar component ($x$ $=$ 0, 1 and 3) to the radiation
field.  The stronger the disk radiation field the more polar the flow 
becomes.  Adding in an increasingly large stellar component causes the 
outflow to become more equatorial.
}
\end{figure*}

Our models for $\alpha = 0.4$ and $0.8$ indicate that the wind mass loss
rate is very sensitive to the rapidity with which the force multiplier
saturates at its maximum value (i.e. the limit in which all lines have
become optically-thin).  For $\alpha=0.8$, $\MDOT_w$ 
is $\sim 4$ orders of magnitude higher than for $\alpha=0.4$.  Higher 
$\alpha$ yielding higher mass loss simply reflects the fact that the
force multiplier is higher for a given $t$ (where $t$ is the optical depth 
parameter defined in equation~7).  The extremity of the effect shows that it 
does not require large shifts in the relative magnitudes of the radiation 
force and effective gravity to make the difference between negligible and 
efficient mass loss.   Interestingly, there seems to be no change
in the power law dependence of $\MDOT_w$ on luminosity between $\alpha = 0.6$
and $\alpha = 0.8$, in contrast to the behaviour of one-dimensional
stellar wind solutions.

Figure~9 presents how the wind radial velocity at $10 r_\ast$ changes
with $\MDOT_a$ and  $[1+x] L_D M_{max}$ for various $\alpha$.  The figure we 
quote is $v_r$ of the gas at a representative angle in the supersonic part of 
the outflow -- this is usually $v_r$ at the polar angle where $\rho v_r$ 
peaks in the fast stream (see Figure~3).  Generally, the wind velocity 
is a weaker function of the disk luminosity than $\MDOT_w$.  Figure~9b 
suggests that $v_r$ increases with $[1+x] L_D M_{max}$ along one universal 
curve at fixed $\alpha$, mirroring the single relation found also for the 
mass loss rate.  This curve is not as smooth as that for $\MDOT_w$ because of 
inherent imprecision in our method of determining $v_r$.  The radial velocity 
is a strong function of $\theta$ and also may change with time.  To obtain a
smoother curve we would have had to calculate many of our models for longer 
and then derive a consistent set of time averages rather than make `by eye'
measurements as here.  Despite this, the trend is clear enough that
$v_r$ scales with $[1+x] L_D M_{max}$ in our wind models.

While it is true of our models that the integrated mass loss rate and
typical radial outflow velocity are not sensitive to the particular
geometry of the radiation field, this does not mean that the radiation
geometry has no role to play.  Panels b, c and d of Figure 10
compare the flow pattern from three models in which the mass accretion
rate (and therefore disk luminosity) is held fixed at $\MDOT_a = \pi
\times 10^{-8}$ M$_{\odot}$ yr$^{-1}$ and the luminosity of the central
star is varied using $x=0, 1,$ and 3.  Note this implies that the total
luminosity (disk plus star) is increasing.  It can be seen that, as the
contribution of the central star to the radiation field grows, the flow
becomes more equatorial.  This is unsurprising given that the
increasing contribution from the central star boosts the radial
component of the radiation force, while contributing little to the
(negative) $\theta$ component.  

However, it is important to realise that the flow geometry also responds, for 
a fixed radiation geometry, to a change in the driving luminosity.  Panels a 
and b of Figure 10 compare the flow pattern from two models in which the mass 
accretion rate (and therefore disk luminosity) is changed from $\MDOT_a =
10^{-8}$ M$_{\odot}$ yr$^{-1}$ to $\MDOT_a = \pi \times 10^{-8}$
M$_{\odot}$ yr$^{-1}$, while the central star is assumed to be dark ,
i.e. $x=0$.  The contrast between these two panels shows how a more
luminous disk will power a stronger, more vertically-directed wind.
Specifically, an increase of a factor $\pi$ in the mass accretion rate
is sufficient to divert the fast boundary stream from $\theta \sim
45^o$ to $\theta \sim 30^o$.

\subsection{Sensitivity of models to assumptions}

There are two parameters in our models listed in Table 1 whose values
seem arbitrary, yet which may appear important in determining our
solutions.  These are the density at the base of the wind $\rho_0$
and the maximum value of the force multiplier $M_{max}$.
In this section, we discuss the effect on our models of varying these
parameters.  We also discuss the tests we have performed of the
sensitivity of our solutions to the approximations made in 
calculating the line-driving.

First, we consider the effect of varying $M_{max}$.  In principle,
$M_{max}$ is a function of $k$, $\alpha$ and $\eta_{max}$ (cf. equation 6).
In this paper, we have treated $k$ and $\alpha$ as free parameters and
studied the effect of varying $\alpha$ in section 4.3.  In this case,
$\eta_{max}$ was also varied to keep $M_{max}$ fixed.  Instead, here
we investigate the effect of varying $\eta_{max}$ (and therefore $M_{max}$)
for fixed $k$ and $\alpha$.  As pointed out in section 2, $M_{max}$
is not an arbitrary quantity, but we expect it only to take effect in regions 
where the optical depth $t$ is small.  Since $\MDOT_W$ is likely to be fixed 
near the wind base in the higher $t$ domain, we may anticipate that 
moderate changes in $\eta_{max}$ ($M_{max}$) are more likely to have a 
bearing on the strongly supersonic flow and alter, for example, the wind 
terminal velocity.  Our tests indicate this is the case.  Of course very low 
$\eta_{max}$ ($M_{max}$) can adversely affect the driving near the base of 
the flow and, if low enough, quench the wind altogether.  If we treat 
$M_{max}$ as the parameter identifying the scale of the line driving force, 
we can use it to set the following rough lower limit on the total luminosity 
needed to produce a fast disk wind:
\begin{equation}
(1 + x) L_D \gtappeq  L_{Edd}/M_{max} ,
\end{equation}
where $L_{Edd}= \frac{4\pi c G}{\sigma_e} M_\ast$ is the Eddington
luminosity.  

Next, we consider the effect of varying the density $\rho_0$ at the
base of the wind.  The radiation force due to lines per unit mass is a
function of gas density (equations 2--4) such that the higher the density, 
the lower the force.   For $\theta$ near $90^o$, the total force acting on
the gas is nearly equal to the radiation force since the effective
gravity near the disk mid-plane is small.  In this region, therefore,
the boundary density $\rho_{0}$ controls not only the radiation line
force but also the total force acting on the gas.  We have tested the
sensitivity of our model winds to changes in our assumed value for
$\rho_{0}$.  We find that for $\rho_{0} \gtappeq 10^{-10}~{\rm g~cm^{-3}}$,
the flow is transsonic (i.e. the subsonic portion of the
flow is resolved on our numerical mesh), and the properties of the
outflow do not depend on the value of $\rho_{0}$.  On the other hand,
for $\rho_{0}$ $< 10^{-10}~{\rm g~cm^{-3}}$ the line radiation force 
per unit mass is high at the outset, giving rise to a steady
supersonic outflow even at the wind base (i.e. the flow becomes
supersonic in less than one grid point).  This is clearly unphysical:
if the radiation force per unit mass were this high in a real disk,
subsonic outflow would have begun at much higher densities deeper
in the disk.

The highest allowed value for the density at the base of the wind must
be less than the density at the midplane of the disk.  In principle,
the density along the disk plane can be determined self-consistently
from disk structure models.  At the same time, the density at the base
of the wind must be large enough to produce a transsonic wind.  Disk
structure models have shown that the density in the disk midplane is
$\gtappeq 10^{-9}~{\rm g~cm^{-3}}$ (e.g. Pringle 1981, Carroll \etal 1985), 
thus the value $\rho_{0} = 10^{-9}~{\rm g~cm^{-3}}$ 
we adopt in our models can be seen to 
be entirely satisfactory.

Finally, we have also examined the sensitivity of our solutions to the
assumptions we adopt to compute the line driving force (see Appendix
C).  While it is infeasible to evaluate the generalized CAK force in our 
models at every timestep,  it is feasible to evaluate it at a particular time 
for all locations in the flow, in order to compare the exact calculation with 
the force computed approximately.  Typically, we find the biggest 
discrepancies very close to the disk surface: the full treatment of the 
generalized force yields an acceleration up to an order of magnitude higher 
than that given by our approximation.  This is because of the extra 
contributions from terms depending on $v_{\phi}$ in the rate of strain tensor 
which we drop.  However, as the flow is accelerated, these terms are quickly 
overwhelmed by terms which depend on $v_{z}$ and $v_{r}$ that are included in 
our approximations.  Thus, a few degrees above the surface of the disk, our
approximate form for the radiation force is in good agreement with the full 
expression.  Moreover, as the optical depth in the lines decreases, the force 
reaches its maximum value $M_{max}$ and becomes independent of the 
approximations we adopt.  Typically the force saturates at just a few
stellar radii in response to the declining wind density.

Still, it is possible that the increased radiation force close to the 
disk plane in a more exact treatment may affect the solutions by, 
e.g., increasing the mass loss rate in the wind for a given disk luminosity.  
Thus, the development of an efficient computational scheme that can relax to a
hydrodynamical solution consistent with the full form of the generalized
line-driving force is important: we will present such results in a
future communication.  However, the current tests of our approximations
give no indication that the key features of our results (the unsteady
nature of radiation driven disk winds, or the overall two-dimensional
geometry of such winds) are sensitive to an improved representation of
the radiation force.

\section{ Discussion }

\subsection{ Origin of unsteady outflow }

The most dramatic result of these models is the discovery of unsteady
outflow in  many of the cases that we have considered.  This component
when present occurs in the base of the outflow near the disk.  It is
characterised by large amplitude density and velocity variations.  It is
important to ask what is the origin of this behaviour.  We identify several
factors which contribute to it.

   The first, and fundamental driver of the behaviour, is the difference in 
the height dependence of the vertical components of gravity and the radiation
force.  The former increases linearly whilst the flux integral central
to the latter is nearly constant in the brightest parts of the disk.  The 
consequence is that mass lifted off the disk plane by radiation 
pressure is susceptible to stalling as the increasing gravity takes effect.  
In this circumstance, mass loss can only be established if a segregation can 
occur in which denser concentrations of mass fall back toward the disk plane, 
while the interspersed lower density gas (in which the line-driving
force per unit mass is larger) continues to be accelerated outward by 
the radiation force.  If this separation were only required to occur in the 
subcritical part of the flow, gas pressure effects might then act to 
smooth the density profile, thereby preventing the development of unstable 
behaviour.  In practice, the radiation force term continues to be at a
disadvantage with respect to gravity out to greater heights in all our
models where the disk is the only source of radiation.  

    A critical aspect that facilitates the unsteady behaviour is
the multi-dimensional character of the flow.  In one dimension, it
is likely that the increase in gravity with height would prevent an
outflow being established at all in the case of a sub-Eddington disk. 
However, in two dimensions, streamlines can merge laterally, with the
result that higher density regions, in which the radiation force per unit mass 
and acceleration is reduced, are created alongside lower density gas that can 
be more readily accelerated to form the outflow.   The contrast between a 
nearly planar flow from a disk and a spherical flow from a star is relevant 
here; the effects of streamline convergence would be reduced by 
geometric dilution in the latter.

    We are certainly not the first to appreciate the significance of the 
increase in effective gravity with height for disk winds.  For example, in 
their essentially one-dimensional treatment, Vitello \& Shlosman (1988) dealt 
with the problem by deriving an ionization structure for the wind which 
ensured the radiation force tracked the rising gravity term.  In contrast, we 
take the view that unsteady behaviour is likely to be a natural characteristic 
of disk winds and therefore see no need to condition our calculations to 
eliminate it.  

   Given the inherent instability present in the outflow, it is not
surprising that our models show complex behaviour.  All that is
required to excite such behaviour are modest perturbations.  These will
arise in our models for several reasons related to the physics of the
problem.  For example, our initial conditions are not a perfect
equilibrium state -- rather, small radial pressure gradients excite
both radial and vertical oscillations of the disk that can seed
perturbations in the outflow.  Vertical oscillations of the disk
continue to be driven as dense material falls back onto the disk from
the flow.  In fact, the tendency of pressure-supported disks to undergo
vertical oscillations (e.g., Cox, \& Everson 1980; Lin, Papaloizou,\&
Savonije 1990) may ensure the flow will never reach a steady state.
In addition to these small amplitude perturbations associated with
the lack of perfect hydrostatic equilibrium in the initial state, there are
large amplitude velocity perturbations associated with the
transients generated during the establishment of the outflow.
Finally, there is considerable velocity shear between the dense disk wind and 
the lower density fast stream defining the upper envelope of the flow. There 
is evidence in our simulations that this shear gives rise to Kelvin-Helmholtz 
instabilities.

   It is well-known that even 1D radiation-driven stellar winds are 
subject to powerful instabilities (OCR).  It is plausible this
instability will be present in radiation-driven disk winds also,
although the instability tends to produce strong shocks perpendicular
to the outflow which we do not observe in our simulations.
Even without these (as described above) there are other physical effects that 
will in any case lead to complex, unsteady flow.

   A related and important feature of our calculations is that the
addition of a strong radial component to the radiation field
associated with a bright central star `organizes' the flow into a
steady state.  The effect is almost certainly caused by the fact that
the streamlines near the surface of the disk will be directed outwards of
the purely vertical by the added stellar radiation.  Thus, the effective 
gravity along the streamlines no longer increases, and the mechanism of the 
unsteady behavior (that gravity exceeds the radiation force at some distance 
from the disk) no longer operates.  Empirically, we have seen that,
in the luminosity domain where the disk wind is robust,
a CS half as luminous as the disk ($x~\simeq~0.5$) is sufficient to make this 
difference.

\subsection{ Application to CV }

   Our present calculations have been motivated by and designed for the
case of winds from CV.  We now consider whether the dynamical structures
and mass loss rates predicted by them are likely to be appropriate.

  The primary evidence for the existence of winds in CVs is contained
within ultraviolet observations of high-state non-magnetic systems
(dwarf novae in outburst and nova-like variables).  In low inclination
non-eclipsing systems, the profiles of the stronger resonance lines
include broad blueshifted absorption indicating outflow.  The 
maximum expansion velocities inferred are on the order of a few thousand 
km~s$^{-1}$ and are thus comparable with the typical white dwarf
escape velocity.  A point of contrast between the line profile
shapes seen in OB stars and CVs, is that deepest absorption is achieved
near terminal velocity in the former, but near line centre in
the latter (e.g. see data presented by Prinja \& Rosen 1995).  In
high inclination eclipsing systems, the P Cygni absorption is replaced by 
broad high contrast line emission.  
The order of magnitude decrease in expansion speed with respect to a
spherically-symmetric MS star wind ($v_{\infty} \sim 1500$ km s$^{-1}$;
Howarth \& Prinja 1989) is particularly significant.  Associated with
the changed mass flux and the restricted opening angle of the outflow
is a density that can be up to $\sim$100 times higher than would
be expected of a spherically-symmetric stellar wind -- the efficiency
of H{\sc i} line emission would presumably rise by a still larger factor.
All of these effects are substantive changes in the right direction,
suggesting that a radiation-driven disk wind model for massive YSOs is
worth further investigation.

   Adopting a mass accretion rate consistent with values inferred from 
observations of high-state CV (i.e. $\dot{M}_{a} \sim 3 \times 10^{-8}$ 
M$_{\odot}$ yr$^{-1}$, e.g. Warner 1987) leads us to consider a qualitative 
comparison of model 3 with observation.  This model is characterised by 
complex dense flow near the equatorial plane bounded by a fast stream (see 
Figure 10a).  Typical velocities in these two components are $\sim200$ km~ 
s$^{-1}$ and $\sim2000$ km~s$^{-1}$ respectively.  On viewing such an object 
at low inclination, we would expect to see high-velocity blueshifted 
absorption due to the fast stream combined with a substantial low-velocity 
absorption component originating in the more slowly churning equatorial gas.  
At high inclinations the low-velocity component should still be apparent in 
absorption, while the high velocity gas will appear in emission if it is no 
longer seen in projection against the bright inner disk.  Hence both at
low inclination and at high inclination, the kinematic structure of the
model outflow appears to be capable of matching the characteristics indicated 
by observation.  

   Clearly, it will be appropriate to confirm qualitative impression of
agreement by carrying out detailed line profile synthesis based on these 
models.  When this is undertaken, it may well be appropriate to think again 
about the boundary condition currently imposed at the surface of the white 
dwarf.  In the interests of simplicity we have thus far ignored the possibility
that there may yet be a significant component of boundary layer emission
between the disk and star, and have not allowed any mass loss from the star
itself.  In a crude way, the $x = 1$ models give some idea as to what
impact the presence of a hot white dwarf and non-planar boundary layer might
have.  Mass loss from the star could very well add significantly to the 
total column contained within the fast stream -- particularly if the star
is allowed to rotate at a significant fraction of break-up.  These are,
however, issues that amount to the introduction of further free parameters
that should be faced in the future, rather than taken on board now, at the 
outset.

   If, as our models suggest, there is an equatorial zone of complex
time-dependent flow, there are consequences of this that may be directly
observable.  We find the flow varies on timescales of order of the local 
orbital period, i.e. a few tens of seconds in the vicinity of a white dwarf.  
If this behaviour is present in real systems and gives rise to a granularity
on a spatial scale not too small compared with the total extent of the 
effective resonance line-forming region, we can expect the low velocity 
absorption component to vary on this timescale.  This prediction is just 
within the realms of testability using highly time resolved HST spectra.  
The effect may be looked for both in high and low inclination systems

   We now come to the question of the comparison between model and observed
mass loss rates.  We find for $\dot{M}_a \sim 3 \times 10^{-8}$ M$_{\odot}$ 
yr$^{-1}$ that the mass loss rate in the wind is $5 \times 10^{-12}$ 
M$_{\odot}$ yr$^{-1}$ for $\alpha = 0.6$, rising to almost
$1 \times 10^{-10}$ M$_{\odot}$ yr$^{-1}$ for $\alpha = 0.8$ (see
Table 2 and Figure 8).  The reason for this sensitivity is that the higher
value of $\alpha$ causes the force multiplier to achieve its maximum value
earlier in the flow.  Observational lower limits based on profile fitting 
uncorrected for unknown ion abundances span much the same range (e.g. Drew 
1997, Prinja \& Rosen 1995).  Estimates based on ionization models require 
$\dot{M}/\dot{M}_a$ in the region of a few percent (Hoare \& Drew 1993).  In 
view of our expectation that the present calculations are liable to 
underestimate the wind mass loss, this initial comparison is very encouraging 
indeed.  However, it is also true that there is yet much work to be done to 
determine internally consistent choices for the parameters $k$ and $\alpha$  
that control the radiation force multiplier.  Thus far we have just used 
values typical for single hot stars.   

   Lastly, we note that the mass loss is modelled as showing a sharp cut-off 
as the product of the total luminosity and maximum force multiplier decreases 
below a critical value.  Presently this is twice the Eddington limit,
and translates at small $x$ and $\alpha = 0.6$ into $\dot{M}_a \simeq 
10^{-8}$~\MSUNYR.  There is a parallel to this behaviour in 
ultraviolet observations of dwarf novae undergoing outburst, where it has 
been noted that P-Cygni absorption features are apt to disappear 
very suddenly as the decline from maximum light begins.  A good example of
this was seen early in a decline of SU~UMa (Woods, Drew \& Verbunt 1990), when
a factor of 2 decrease in the UV continuum erased what had been prominent 
blueshifted absorption in C{\sc iv}~1549\AA\ and other lines at maximum.
Another aspect of this is that different systems apparently present very 
different levels of mass loss, despite the expectation that the high state
viscosity, and hence mass accretion rates, cannot vary by more than a factor
of a few (e.g. compare and contrast the weak blueshifted absorption features
in SS~Cyg, during outburst, with the extremely strong features in RW Sex, a 
nova-like variable, Prinja \& Rosen 1995).  Ultimately this effect will
provide a useful quantitative calibration of radiation driven disk wind 
models against observation.  For the timebeing, it is again encouraging that 
the cut-off occurs at a mass accretion rate comparable with those believed 
to be attained during outburst.

\subsection{ General scaling }

   The models presented in this paper have all been calculated for white
dwarf accretion disks.  We show below how our models might be scaled to
produce guideline mass loss rates and expansion velocities for other
applications.

   We introduce a set of primed dimensionless variables.  First, it is
natural to scale lengths to the stellar radius, $r_\ast$:
\begin{equation}
	r = r' r_\ast
\end{equation}
and define the unit time $\tau = \sqrt\frac{r_\ast^3}{GM_\ast}$, as earlier,
such that
\begin{equation}
	t = t' \tau .
\end{equation}
The unit velocity is accordingly $v_o =\sqrt\frac{GM}{r_\ast}$.  For the
white dwarf case this is $3017 ~\rm km~s^{-1}$.  Translational velocity
and the sound speed then become: 
\begin{equation}
	{\bf v} = {\bf v'} v_o
\end{equation}
and
\begin{equation}
		c_s = c'_s v_o
\end{equation}
In our models, $c'_s=4.6\times10^{-3}$.
The Eddington factor expressed in terms of just the disk luminosity is
\begin{equation}
	\Gamma=\frac{\sigma_e \MDOT_a}{8\pi c r_\ast }
\end{equation}
where $\sigma_e$ is the Thompson scattering cross-section divided by the
mass of the hydrogen atom.

    Using these new variables, the equation of motion can be rewritten in 
the dimensionless form:
\begin{eqnarray}
   \rho' \frac{D{\bf v'}}{Dt'} & = & - {c'_s}^2 \nabla (\rho' ) + 
   \rho' \frac{1}{{r'}^2} + \nonumber \\
 &  &  \frac{6\rho'\Gamma}{\pi} \left({\bf f}_{D}(x)+{\bf f}_{D}^{l}(x,M)+
  ~{\bf f}_{\ast}(x)+{\bf f}_{\ast}^l(x,M) \right), \nonumber \\
 &  & 
\end{eqnarray}
where the scaling to a dimensionless density via $\rho = \rho' \rho_o$
is trivial.
This equation has three parameters: $c'_s$, $\Gamma$, $x$, and depends on one 
dimensionless function -- the locally-determined force multiplier, $M(t)$.  
We can approximate and hence simplify this somewhat.  First, for many cases 
of interest, the 
electron scattering terms will be of minor importance compared to the line 
acceleration terms and so may be neglected.  Second, we may conclude from the
empirical absence of a dependence upon $x$ in the relations between total 
luminosity and either mass loss rate or outflow velocity (Figs. 8b \& 9b)
that the disk and CS driving terms can be combined to yield:
\begin{equation}
   \rho' \frac{D{\bf v'}}{Dt'} \simeq - {c'_s}^2 \nabla (\rho' ) + \rho' \frac{1}{{r'}^2}
 + \frac{6\rho'}{\pi}(1 + x)\Gamma M(t) f'(r,\theta)
\end{equation}
wherein $f'(r,\theta)$ is a factor encompassing all the geometric aspects of 
the radiation force calculation.  A priori it was not possible to assume that 
the dynamics might be reducible to such a form.  

   It only remains to provide a scaling to allow mass loss rates to be
estimated for other applications.  This can be extracted from the definition
of the dimensionless Eddington factor, $\Gamma$, in that we can define a 
fiducial mass time derivative such that $\MDOT_{o} = 8\pi c 
r_{\ast}/\sigma_{e}$.  The mass loss rate will then scale as
\begin{equation}
   \MDOT_{w} = \MDOT_{w}' \MDOT_{o}
\end{equation}
In Figs. 8b and 9b we provide as alternate ordinates the quantities 
$\MDOT_{w}'$ and $v_{r}'$ in order to facilitate rescaling of our results
to other contexts for which $\MDOT_o$ and $v_o$ can be estimated.

\subsection{Other astrophysical applications}

Although the models presented in this paper have been motivated
primarily by observations of CVs, there are clearly other astrophysical 
systems to which our results may be relevant.  Here we discuss just
two such cases: accretion disks associated with active galactic nuclei (AGN), 
and massive young stellar objects.

The presence of broad, blueshifted absorption lines in quasar spectra
(Osterbrock 1989) is often interpreted as evidence for a line-driven
disk wind.  Recently, Murray et al (1995) have constructed dynamical
models for such winds based on the solution of the one-dimensional
(radial) equation of motion subject to certain assumptions about how
the gas is loaded onto radial streamlines via vertical motions.  It is
not clear whether these assumptions lead to a good representation of
the streamlines in a fully two-dimensional solution such as presented
here.  

The bulk of the radiative flux in quasars comes from or near the
central source, implying our models with very large values for the
parameter $x$ should be most appropriate to these systems.  The most extreme
value of $x$ we have considered is 10 (model 14).  In it, we find strongly 
radial flow confined to angles of less than 30 degrees from 
the disk midplane with little time-dependence.
However, the radiation from the central source in the AGN case is very much 
harder than that produced locally in the disk and, as the former increasingly 
dominates over the latter with increasing height above the disk photosphere, 
it is plausible that the force multiplier would become a function of position 
to reflect this (see Vitello \& Shlosman 1988, Murray et al. 1995).
This is not an effect that our present models include.  At least a 
simplified treatment of the photoionization and recombination of the wind 
material is required before the two-dimensional structure of quasar winds 
can be examined self-consistently.

In the case of high mass young stellar objects, e.g. the BN-type objects and 
Herbig Be stars, photoionization effects are not an overriding concern in that
the literature already contains force multiplier parameters designed for
the appropriate effective temperature range.  A more fundamental issue is the 
nature and extent of their circumstellar disks, as this cannot yet be said to 
have been defined compellingly.  That disks of some kind are present has been 
entertained by many (Simon et al. 1985; Hamann \& Persson 1989; Chandler, 
Carlstrom \& Scoville 1995 -- to mention a few).  A major phenomenological 
challenge of these objects is the dynamical origin of their often extremely 
bright, yet modestly velocity-broadened ($\Delta v_{FWHM} \sim 200$ 
km~s$^{-1}$) hydrogen line emission.  If, like classical T~Tau stars, these 
objects are in an active accretion phase, the ratio of stellar to disk 
luminosity may not be too extreme.  For instance, an early B star accreting 
at a rate of $\sim10^{-6}$ M$_{\odot}$~yr$^{-1}$ would be described by 
$x \sim 100$.  Since, for $x \gtappeq 10$, the disk's light is dominated
by the reprocessed component, there is no susbtantive difference between
$x = 10$ and any higher value of $x$. Thus, our model 14 with $x = 10$ may 
again be crudely indicative of the outflow geometry we might expect for such 
systems.  The expectation is therefore that the outflow would be equatorial 
and steady.  

A more interesting point, however, is that the flow is very likely to be very 
much more dense and significantly less rapidly expanding than a conventional 
early-type stellar wind.  Specifically, the effective Eddington number 
($M_{max}L/L_{edd}$) for an early B star is likely to be in the region of 
20 or so, while the scaling variables, $v_o$ and $\MDOT_o$ (section 5.3), are 
respectively $\sim500$ km s$^{-1}$ and $\sim0.01$ \MSUNYR .  These numbers 
combine with the results in Figs. 8b and 9b to yield mass loss rates estimates
in excess of 10$^{-8}$ \MSUNYR and maximum expansion velocities of 
$\sim500$ km s$^{-1}$ (i.e. $v_r' \sim 1$).  This amounts to an order of 
magnitude increase in $\MDOT_w$ and a factor of a few decrease in 
expansion speeds with respect to a spherically symmetric MS star wind 
($\MDOT _w \sim 10^{-9}$ M$_{\odot}$~yr$^{-1}$, $v_{\infty} \sim 2000$ km 
s$^{-1}$; Howarth \& Prinja 1989).  The net impact of both these differences 
and the restricted opening angle of the outflow could be to raise the density,
with respect to a normal MS stellar wind, by a factor of a few tens
and the efficiency of H~{\sc i} line emission by a factor of 100--1000 
perhaps.  All of these effects are substantive changes in the right 
direction, suggesting that a radiation-driven disk wind model for massive 
YSOs is worthy of further investigation.

\subsection{Limitations of the present models}

There are a number of limitations of the present analysis which are
worthy of mention and further investigation.  

Perhaps the most important relate to the approximations adopted here to
represent the radiation force.  We have already discussed, in section
4.4, the tests we have performed to check the sensitivity of our models
to an improved representation of the general line-driving force in a
multidimensional wind.  Based on these tests, we conclude it is
unlikely that the major results of this paper (for example, the
two-dimensional geometry of line-driven winds from disks, or the
existence of unsteady behavior in low luminosity systems) will change
with a formalism which includes all terms in the radiation force on
lines.  However, quantities such as the mass loss rate and terminal
velocity reported here should only be considered accurate to factors of
a few.  Here, we also wish to point out that in a rotating wind there
are azimuthal forces even in axisymmetry (because not all terms in the
velocity gradient projected along the line of sight $dv_l/dl$ are
symmetric in $\phi$, see equations 8 and A2); these forces may change
the angular momentum of the gas and effect the dynamics of the wind.
While we expect such effects to be small, we have yet to study them in
detail.  We shall report the results of our calculations using a more
general treatment of the radiation force on lines in a two-dimensional,
rotating wind in a future communication.

Of course, a more fundamental concern is whether the Sobolev
approximation should even apply in principle to the multidimensional
and time-dependent flows considered here.  In adopting the Sobolev
approximation, we have ignored non-local radiative transfer
effects.  Because the velocity field in some of the models reported
here is neither monotonic nor steady, non-local effects such as
shadowing can be expected to affect the solutions.  A proper study of
these effect requires the use of algorithms for multidimensional transport of
line radiation in a rotating wind, which is beyond the scope
of the present work.  However, it may be anticipated that the 
inclusion of shadowing would have a similar effect to increased $x$
for the reason that shadowing should mostly reduce the driving of
the slow equatorial component and have little impact on the 
relatively well-organized fast boundary stream.

Other effects which bear further investigation are the inclusion of
mass loss from the central star.  This however requires a realistic
prescription for the properties of the radiation field and gas flow in
the interaction region (boundary layer) between the CS and accretion
disks.  Since magnetic fields are likely to be central to the
production of angular momentum transport in accretion disks (Balbus \&
Hawley 1997), it would also be fruitful to consider the effect of a global
magnetic field anchored in the disk on the properties of the wind.

Finally, we have considered the two-dimensional structure of winds
assuming an isothermal equation of state.  Thermal pressure effects can
be expected to be important only in the subsonic acceleration zone,
which we find is generally small in spatial extent.  Nevertheless, we
have not modeled the transition between the optically-thick (and therefore
adiabatic) gas inside the disk, and the optically thin wind above.  
In principle, the radial variation in the internal structure of the disk 
caused by the radial variation in temperature might affect conditions at the 
base of the wind. This inadequacy is not so serious given that the main seat 
of the outflow is the relatively small area of the innermost disk 
($r \ltappeq 2r_{\ast}$).

Clearly dynamical models which consider the internal magnetohydrodynamics of 
an optically thick, turbulent accretion disk (Brandenburg et al 1995; Stone 
et al 1996) and radiation pressure on spectral lines in the wind region above
the disk are the most appropriate description of real disks; such
models await future studies.

\section{ Conclusions }

Using numerical methods to solve the two-dimensional, time-dependent
equations of hydrodynamics, we have studied radiation driven winds from
luminous accretion disks.  In so doing we have accounted for the
radiation force mediated by spectral lines using a generalized
multidimensional formulation of the Sobolev approximation. Our primary
conclusions are the following.

(1) We find radiation driven winds from luminous accretion disks
are intrinsically unsteady: the outflow consists of large
amplitude density and velocity fluctuations, with some regions of
dense material undergoing infall.  This behavior
is rooted in the difference in the variation with height of the
vertical component of gravity and the radiation force.  Since the former
increases, it grows until it overwhelms the radiation force,
causing high density material (in which the radiation force per unit mass is
low) to stall.  Despite the fact that instantaneous values in the wind
are variable, time-averaged values are constant.

(2)  The contribution of a strong radial component to the driving
radiation field from a bright central star serves to `organize'
the outflow into a steady state.  Very bright central stars
produce steady transsonic disk winds.  Moreover, the region producing 
unsteady outflow is reduced as the luminosity of the disk is increased.

(3)  Regardless of whether the flow is steady or unsteady, we find the
time-averaged geometry of the flow typically consists of a dense, nearly
equatorial, low velocity flow confined to angles within $30^o$ to $45^o$ of 
the equatorial plane, bounded by a lower density, high velocity flow in a 
channel at larger angles.  In the absence of a wind directly from the 
central, star the gas density in the polar regions is so low as to be of no
dynamical or observable significance.  Most of the mass
loss occurs within a few stellar radii of the central star.

(4)  The geometry of the radiation field is a major factor in controlling
the geometry of the outflow.  Increasing the luminosity of the star
at a fixed disk luminosity produces a radial wind confined to smaller
regions near the equatorial plane.  Conversely, increasing the disk
luminosity at a fixed stellar luminosity produces a more polar outflow.

(5)  The total mass loss rate and terminal velocity in the wind
depends on the total luminosity of the star plus disk system, but
is insensitive to the outflow geometry or whether the wind is steady or
unsteady.  No outflow is produced if the effective luminosity
of the disk (that is, the luminosity of the disk times the maximum
value of the force multiplier associated with the line-driving force)
is less than the Eddington limit.  Above the Eddington limit,
the mass loss rate in the wind scales with the effective luminosity as a
power law with index of about 1.5.  The effective luminosity can
be increased either by increasing the accretion rate in the disk,
or by increasing the brightness of the central star.
The ratio of the mass loss rate in the wind to the accretion rate
increases sharply, reaching a few percent for the most luminous disks
considered.

This study has been motivated primarily by high resolution spectroscopic 
observations of winds from disks in CV systems.  The overall structure of 
disk winds revealed by our calculations, i.e. a dense equatorial wind with a 
fast polar outflow, appear to be in agreement with the kinematics inferred 
for real systems.  Furthermore, the magnitude of the mass loss rates obtained 
on adopting a force multiplier parameterisation known to be applicable
to OB stars overlaps the range that has been deduced from observation.
We plan more detailed comparison of line profiles computed from our models 
with observational data in the future.  Future applications also include high 
mass YSOs with circumstellar disks (in which case outflow from the
central star must also be allowed), and AGN (in which case
photoionization of the wind by the central source must be taken into
account).

{\bf Acknowledgments:} This research has been supported by a research
grant from PPARC, and by NASA through HST grant GO-6494.  Computations
were performed at the Pittsburgh Supercomputing Center.

\onecolumn
\appendix

\section{Calculation of the radiation field from disk and central star}

We use a spherical polar coordinate system with an origin at point C, the 
center of the CS.  Colatitude ($\theta$) is measured from the rotation axis 
of the disk, and azimuth ($\phi$) is measured from a plane perpendicular to 
the disk plane, containing the point C and a point W above the disk (see 
Figure~A.1).  We define the location of a wind point, W, and  a disk point, D,
by the co-ordinates $(r,\theta, 0^o)$ and $(r_D, 90^o, \phi_D)$
respectively.  The distance 
between D and W then is 
\begin{equation}
d_D = (r_D^2 + r^2 - 2 r_D r~\cos~\beta_D)^{1/2},
\end{equation}
where $\beta_D$ is the angle WCD and $\cos~\beta_D =  \sin~\theta~\cos~\phi_D$.
The direction D toward W can be defined by the unit vector 
${\hat{n}}~=~(n_r, n_\theta, n_\phi)$. Using the coordinates of points D 
and W,
\begin{equation}
 n_r~=~\frac{r - r_D~\sin~\theta~\cos~\phi_D}{ d_D};~~ 
 n_\theta~=~\frac{r_D~\cos~\theta~\cos~\phi_D}{ d_D};~~ 
 n_\phi~=~\frac{r_D~\sin~\phi_D}{ d_D}. 
\end{equation}

The intensity of an $\alpha$-disk at point D is (e.g., Pringle 1981)
\begin{equation}
I_D(r_D) = \frac{3 G M_\ast \MDOT_a}{8 \pi^2 r_D^3} 
\left(1 - \left(\frac{r_\ast}{r_D}\right)^{1/2}\right),
\end{equation}
where $M_\ast$  and $r_\ast$ are the mass and radius of the central star, 
$\MDOT_a$ is the accretion rate through the disk (Shakura \& Sunayev 1973). 
The total luminosity of an $\alpha$-disk is
\begin{equation}
L_D~=~\frac{G M_\ast \MDOT_a}{2 r_\ast}.
\end{equation}

In the presence of a luminous CS, the intensity radiated by an 
optically-thick $\alpha$-disk changes due to heating of the disk by the 
CS radiation.  To calculate this illumination effect, it is convenient to 
describe the location of the CS surface point S in a spherical polar 
coordinate system $(R, \Theta, \Phi)$ in which the origin is at the
point D.  The colatitude $\Theta$, is now measured from the DC axis and 
the azimuth $\Phi$, is measured from the plane perpendicular to the disk 
surface that contains both the points D and C.
Expressing the central star luminosity in $L_D$ units 
\begin{equation}
L_\ast~=~x L_D~=~x\frac{G M_\ast \MDOT_a}{2 r_\ast}.
\end{equation}
and assuming that the CS surface is isothermal, the CS intensity then  is 
\begin{equation}
I_\ast~=~\frac{L_\ast}{4\pi^2 r_\ast^2}~=~x\frac{G M_\ast \MDOT_a }{8 \pi^2 r^3_\ast}.
\end{equation}
The stellar energy absorbed per unit time by a surface element of the disk is
\begin{equation}
\frac{dE}{dAdt}~=~I_\ast~\int_{-\pi/2}^{\pi/2}~\int_0^{\Theta_{{max}}} 
\cos\Phi \sin^2\Theta~d\Theta d\Phi~=~I_\ast~(\Theta_{max} -  \sin\Theta_{max} 
\cos\Theta_{max}),
\end{equation} 
where, $\sin \Theta_{max}~=~r_\ast/r$. 

Assuming that the disk reemits the absorbed energy locally  
as a black body, the disk intensity due to irradiation can be written as  
\begin{equation}
I_{out}~=~\frac{I_\ast}{\pi}\left(\arcsin \frac{r_\ast}{r_D} - 
\frac{r_\ast}{r_D} \left(1 - 
\left(\frac{r_\ast}{r_D}\right)^2\right)^{1/2}\right).
\end{equation} 
Thus, using equations A3, A6 and A8,  we can express 
the intensity of the steady state disk illuminated by the CS as 
\begin{equation}
I_{D}(r_D)~= \frac{3 G M_\ast \MDOT_a}{8 \pi^2 r_\ast^3} 
\left(\frac{r_\ast^3}{r_D^3}\left(1 - 
\left(\frac{r_\ast}{r_D}\right)^{1/2}\right)
+\frac{x}{3\pi}\left(\arcsin \frac{r_\ast}{r_D} - 
\frac{r_\ast}{r_D} \left(1 - 
\left(\frac{r_\ast}{r_D}\right)^2\right)^{1/2}\right)\right).
\end{equation}

Finally, at a point W, the radiation flux from a disk surface element  
between ($r_D, r_D+dr_D$) and ($\phi_D, \phi_D+d\phi_D$) is
\begin{equation}
 d{\cal F}_{D}~=~\frac{I_{D}(r_D) ~\cos(DWW')}{d_D^{2}} r_D dr_D d\phi_D,
\end{equation}
where $W'$ is the projection of W on the disk plane 
($ \cos(DWW')  = (r / d_D)~ \cos~\theta $).

\begin{figure*}
\begin{picture}(80,300)
\put(0,0){\includegraphics{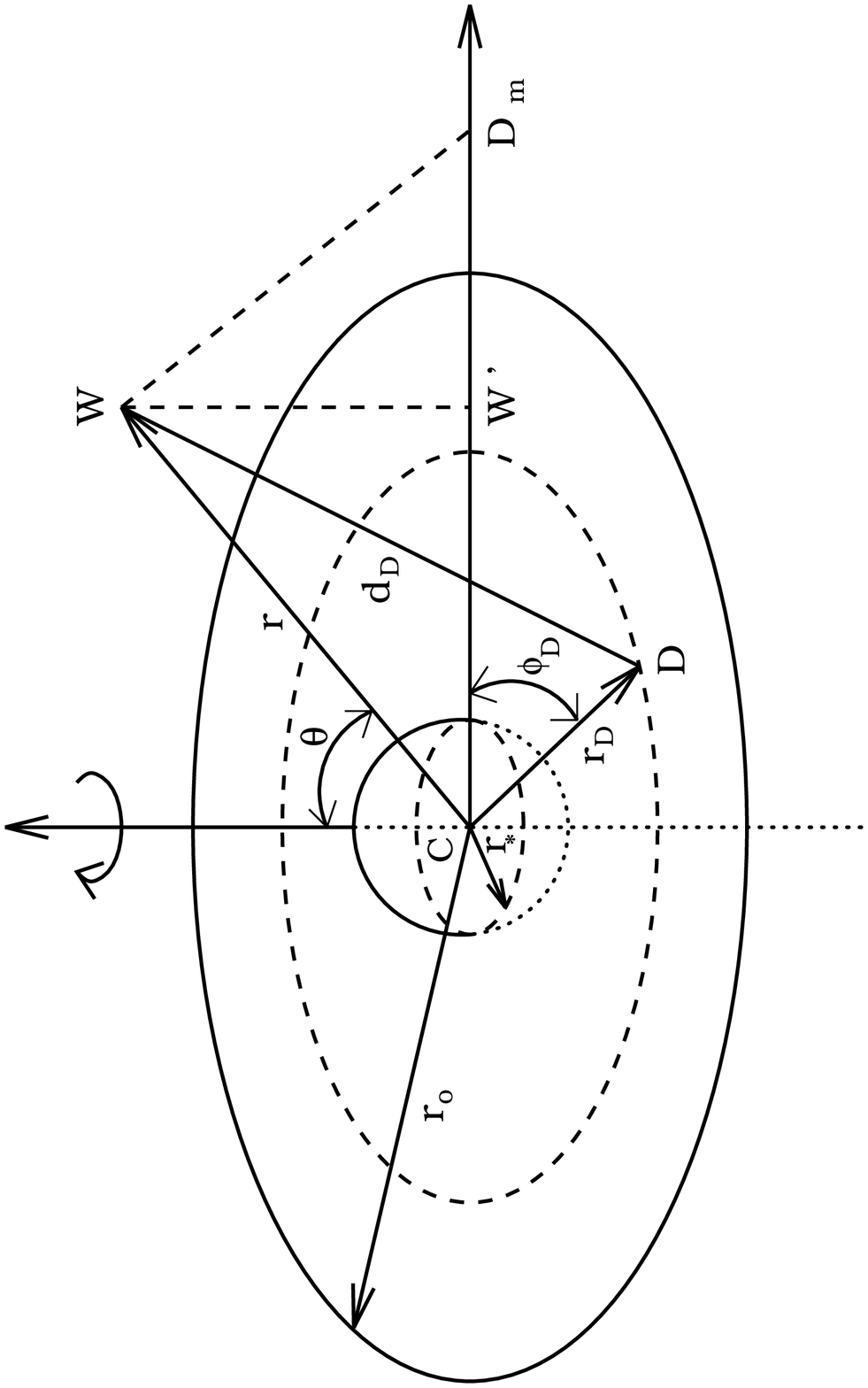}}
\end{picture}
\caption{
A sketch of the disk and central star, identifying the meaning of the
symbols used in the construction of the radiation force integrals.  Note
that the angle $\hat{CWD_m}$ is defined to be $\pi/2$.
}  
\end{figure*}

To calculate the radiation flux from the CS at a point W, we describe 
the location of a point S in a spherical polar coordinate system whose
origin is at the centre of the star (point C).  The angle of colatitude 
$\theta_S$ is measured from the CW axis, while the azimuth $\phi_S$ is 
measured from the plane perpendicular to the disk surface containing the 
points W and C.  The distance between S and W is then
$d_\ast~=~(r^2+r_\ast^2-2rr_\ast\cos\theta_S)^{1/2}$. The unit vector,
$\hat{m}$ specifying the direction SW has the following components:
\begin{equation}
 m_{r}~=~\frac{r - r_\ast~\cos~\theta_s}{ d_\ast};~~
 m_{\theta}~=~\frac{r_\ast~\sin~\theta_s~\cos~\phi_s}{ d_\ast};~~ 
 m_{\phi}~=~\frac{r_\ast~\sin~\theta_s \sin~\phi_s}{ d_\ast}.
\end{equation}

At the point W, the radiation flux due to the stellar surface element 
between ($\theta_s$, $\theta_s+d\theta_s$) and ($\phi_s$, $\phi_s+d\phi_s$) 
is 
\begin{equation}
d{\cal F}_\ast~=~ \frac{I_\ast ~\sin \theta_s 
(r \cos \theta_s - r_\ast)} {d_\ast^3} r_\ast^2
 d\theta_s d\phi_s.
\end{equation}

\section{Calculation of the radiation force from disk and central star}

The radiation force due to electron scattering, per unit mass, contributed by 
the disk surface element along $\hat{n}$ is
\begin{equation}
d{{\bf F}^{rad, e}_D} = {\hat{n}}~
\frac{\sigma_e }{c} d{\cal F}_{D}
\end{equation}
where $\sigma_e$ is the mass scattering coefficient of free electrons.
We assume that the mean mass of the particle is equal to the proton
mass, $m_p$. Thus $\sigma_e~=~\sigma_T/m_p$, where $\sigma_T$ is the Thomson
cross section.
% have we perhaps given ourselves 40 percent or so more force by associating
% a mean mass of just the proton (without a molecular weight factor to give
% a more realistic composition) ????

Using equations A9 and A10 we obtain from equation B1 the radiation force 
per unit mass from the total disk surface acting a point W: namely,
\begin{equation}
{{\bf F}^{rad, e}_{D}} = \frac{3 \sigma_e G}{4\pi^2 
c}\frac{M_\ast\MDOT_a}{r_\ast^3} 
~{\bf f}_{D}({r'},\theta)
\end{equation}
where ${\bf f}_{D}$ is the vector-valued integral
\begin{eqnarray}
{\bf f}_{D}({r'},\theta) & = &\int_{\phi_l}^{\phi_u} \int_{r'_i}^{r'_0} 
\hat{n} \frac{r'~\cos~\theta} {{r'_D}^2 {d'_D}^3} \nonumber \\
& &
\left(1- \left(\frac{1}{{r'}_D}\right)^{1/2}
+\frac{x {r'_D}^3}{3\pi}\left(\arcsin \frac{1}{r'_D} - \frac{1}{r'_D} \left(1 - 
\left(\frac{1}{r'_D}\right)^2\right)^{1/2}\right)\right)  dr'_D d\phi_D, 
\end{eqnarray}
in which the primed quantities are expressed in $r_\ast$ units.
The integration limits,
$r'_i$, $\phi_l$, and $\phi_u$ are functions of position 
because of the need to account for the shadowing of the disk by the CS.
The upper limit on the radial integration is always the outer radius of
the disk, $r'_0$.  The integral with respect to $\phi$ is calculated assuming
symmetry about the $\phi = 0$ plane.
% ...last sentence to explain the factor of two...

Using the CAK formalism (see \S2) the radiation force per unit mass, due to 
spectral lines along $\hat{n}$, contributed by a disk surface element is
\begin{equation}
d{{\bf F}^{rad, l}_D} = {\hat{n}}~
\frac{\sigma_e }{c} d{\cal F}_{D} M(t),
\end{equation}
where $M(t)$ is the force multiplier.  Following the analogy with 
$F_{Di}^{rad,e}$, the total radiation force due to lines from the disk is
\begin{equation}
{\bf F}^{rad, l}_{D} = 
\frac{3 \sigma_e G}{4\pi^2 c}\frac{M_\ast \MDOT_a}
{r_\ast^3}~{\bf f}_{D}^{l}({r'},\theta,{\bf}),
\end{equation}
where ${\bf f}_{D}^l$ is the vector-valued integral
\begin{equation} 
{\bf f}_{D}^l({r'},\theta)  =  \int_{\phi_l}^{\phi_u} \int_{r'_i}^{r'_o} 
{\hat{n}}~\frac{r'~\cos~\theta} {{r'_D}^2 {d'_D}^3} M(t) 
\left(1- \left(\frac{1}{{r'}_D}\right)^{1/2}
+\frac{x {r'_D}^2}{3\pi}\left(\arcsin \frac{1}{r'_D} - \frac{1}{r'_D} \left(1 - 
\left(\frac{1}{r'_D}\right)^2\right)^{1/2}\right)\right)  
d\phi_D dr'_D.  
\end{equation}

The radiation force due to electron scattering from a stellar surface 
element  is
\begin{equation}
d{\bf F}_\ast^{rad, e} ~=~\hat{m}\frac{ \sigma_e}{c} d{\cal F_\ast}
\end{equation}
and using equations A6 and A12, the angle-integrated force becomes
\begin{equation}
{{\bf F}^{rad, e}_\ast} = \frac{3 \sigma_e G}{4\pi^2 
c}\frac{M_\ast\MDOT_a}{r_\ast^3} 
~{\bf f}_\ast({r, \theta}),
\end{equation}
where ${\bf f}_\ast$ is the vector-valued integral
\begin{equation}
{\bf f}_\ast({r, \theta})~=~x~\int_0^{\theta_u} \int_{\phi_i}^{\phi_o} 
~\hat{m}
\frac{\sin \theta_s (r \cos \theta_s - r_\ast) r_\ast^2} 
{3{d'_\ast}^3} 
 d\theta_s d\phi_s,
\end{equation}
By analogy with equation B2, the stellar component of the radiation force 
due to lines is 
\begin{equation}
{{\bf F}^{rad, l}_\ast} = \frac{3 \sigma_e G}{4\pi^2 
c}\frac{M_\ast\MDOT_a}{r_\ast^3} 
~{\bf f}_\ast^l({r, \theta})
\end{equation}
where
\begin{equation}
{\bf f}_\ast^l({r, \theta})~=~x~\int_0^{\theta_u} \int_{\phi_i}^{\phi_o} 
~\hat{m}
\frac{\sin \theta_s (r \cos \theta_s - r'_\ast) r_\ast^2} 
{3{d'_\ast}^3} M(t) d\theta_s d\phi_s,
\end{equation}

Finally, the total radiation force per unit mass acting on a particle at 
point W is
\begin{equation}
{\bf F}^{rad}~=~{\bf F}_{D}^{rad, e}+{\bf F}_{D}^{rad, l}+
{\bf F}_{\ast}^{rad, e} + {\bf F}_{\ast}^{rad, l}
\end{equation}
or using the vector-valued integrals B3, B6, B9 and B11:
\begin{equation}
{\bf F}^{rad}~=~\frac{3 \sigma_e G}{4\pi^2 c}\frac{M_\ast\MDOT_a}{r_\ast^3} 
\left({\bf f}_{D}+{\bf f}_{D}^{l}+{\bf f}_{\ast}+{\bf f}_{\ast}^l \right).
\end{equation}
% ...have placed $x$s such that they always appear within the expressions
% for functions `f'.

\section{The radiation force due to lines for special cases.}

To calculate  the radiation force from the CS we consider the simple case 
where we assume that $Q$ is dominated by terms associated with the radial 
component of the velocity 
\begin{equation}
Q~=~\frac{dv_r}{dr}n_r^2 + \frac{v_r}{r}(n_\theta^2+n_\phi^2)~=
\quad ~\frac{dv_r}{dr}\mu^2 + \frac{v_r}{r}(1 - \mu^2),
\end{equation}
where $\mu~=~{\hat m}\cdot{\bf v}~=~\frac{r-r_\ast\cos\theta}{d_s}$
(see Rybicki \& Hummer 1983).
If we further assume $\mu=1$, i.e. that the star is a point source, then  
\begin{equation}
Q~=~\frac{dv_r}{dr}. 
\end{equation}
This is exactly the case considered by CAK. As we described in \S2,
an advantage of this approximation is that, in the calculation of the 
radiation line force due to the whole star, we can move a time dependent,  
velocity factor outside the integral in equation B11:
\begin{equation}
{\bf f}_\ast^l({r, \theta})~=~k
\left(\sigma_e \rho v_{th} \left|\frac{dv_r}{dr}\right|^{-1}\right)^{-\alpha}
\left[ \frac{(1+\tau_{max})^{(1-\alpha)}-1} {\tau_{max}^{(1-\alpha)}} \right]
\int_0^{\theta_u} \int_{\phi_i}^{\phi_o} 
~\hat{m}
\frac{\sin \theta_s (r \cos \theta_s - r_\ast) r_\ast^2} 
{3d_\ast^3} 
 d\theta_s d\phi_s.
\end{equation}
Thus we need to calculate the integral only once at the beginning of the 
hydrodynamic calculations and update the radiation line force every time step
only through $\frac{dv_r}{dr}$.  However even in a purely radial wind, 
time-dependent calculations become very costly if we take into account a star 
with a finite disk (for example, see the Appendix of CAK, Pauldrach, Puls, 
Kudritzki 1986, Friend \&Abbott 1986).

We also make use of an analogous major simplification of the disk radiation 
force.  Assuming that the gradient of the velocity along the vertical 
direction is the dominant term, equation (8) reduces to 
\begin{equation}
Q~=~\frac{dv_z}{dz} n_z^2,
\end{equation}
where $n_z~=\frac{r' \cos\theta}{d'_D}$.  Numerically, this form of $Q$ is
constructed from its equivalent form in spherical coordinates:
\begin{equation}
Q~=~
\left(\cos^2\theta \frac{dv_r}{dr} + 
\sin^2\theta \frac{1}{r}\left(v_r+\frac{dv_\theta}{d\theta}\right)+
\sin\theta\cos\theta\left(\frac{v_\theta}{r} - 
\frac{1}{r}\frac{dv_r}{d\theta}\right)\right) \left(\frac{r' \cos\theta}{d'_D}\right)^2.
\end{equation}
In this case, equation B6 can be expressed as
\begin{eqnarray} 
{\bf f}_{D}^l(r, \theta) & = & k 
\left(\sigma_e \rho v_{th} \left|\frac{dv_z}{dz}\right|^{-1}\right)^{-\alpha}
\int_0^{\phi_u} \int_{r_i}^{r_o} {\hat{n}}~
\frac{r'~\cos~\theta} {{r'_D}^2 {d'_D}^3}  
\left(\frac{r' \cos\theta}{d'_D}\right)^{2\alpha} 
\left[ \frac{(1+\tau_{max})^{(1-\alpha)}-1} {\tau_{max}^{(1-\alpha)}} \right] \nonumber \\
 & &   
\left(1- \left(\frac{1}{{r'}_D}\right)^{1/2}
+\frac{x {r'_D}^2}{3\pi}\left(\arcsin \frac{1}{r'_D} - \frac{1}{r'_D} \left(1 - 
\left(\frac{1}{r'_D}\right)^2\right)^{1/2}\right)\right)  
d\phi_D dr'_D .
\end{eqnarray}
In the present form we still find the time-dependent $\tau_{max}$ term 
within the integrand.  Strictly, we can not move the factor within square
brackets in front of the integral because $\tau_{max}$ is dependent on 
line-of-sight and hence position on the disk plane.  Therefore we make an 
approximation that $n_z = 1$ as far as $\tau_{max}$ is concerned. Introducing 
a new variable, 
$\tau'_{max}= \sigma_e \rho v_{th} \left|\frac{dv_z}{dz}\right|^{-1} \eta_{max}$,
equation C6 can be rewritten
\begin{eqnarray} 
{\bf f}_{D}^l(r, \theta) & = & k 
\left(\sigma_e \rho v_{th} \left|\frac{dv_z}{dz}\right|^{-1}\right)^{-\alpha}
\left[ \frac{(1+\tau'_{max})^{(1-\alpha)}-1} {{\tau'}_{max}^{(1-\alpha)}} \right] 
\int_0^{\phi_u} \int_{r_i}^{r_o} {\hat{n}}~
\frac{r'~\cos~\theta} {{r'_D}^2 {d'_D}^3}  
\left(\frac{r' \cos\theta}{d'_D}\right)^{2\alpha}\nonumber \\
& &    
\left(1- \left(\frac{1}{{r'}_D}\right)^{1/2}
+\frac{x {r'_D}^2}{3\pi}\left(\arcsin \frac{1}{r'_D} - \frac{1}{r'_D} \left(1 - 
\left(\frac{1}{r'_D}\right)^2\right)^{1/2}\right)\right)  
d\phi_D dr'_D.    
\end{eqnarray}
Now both the time-dependent factors appear outside the integral.  Note that 
our approximation gives exactly the same result as equation C7 for
$\tau_{max} \rightarrow \infty$ because then the factors in square brackets 
are unity in both equations C6 and C7 (see equation 5).
For $\tau_{max} \rightarrow 0$, equation C7 gives values lower 
than C6 because the integrand in equation C7 is smaller than the integrand 
in equation C6 by a factor $(\frac{r' \cos\theta}{d'_D})^{2\alpha}$. 
However for a given $\eta_{max}$,  $\tau_{max} \rightarrow 0$ occurs when $t$
is very small.  Small $t$ will mainly be associated with regions high above 
the disk where the gas density is low.  Bearing in mind that a foreshortened
disk element at low $\frac{r' \cos\theta}{d'_D}$ contributes at low 
weight compared to an element with $\frac{r' \cos\theta}{d'_D} \sim 1$,
we can see that equation C7 is quite a reasonable approximation of
equation C6 even for $\tau_{max} \rightarrow 0$.

\end{document}